\documentclass[twoside,11pt]{article}

\usepackage{blindtext}

\usepackage{jmlr2e}

\usepackage{amsmath}
\usepackage{bm}
\usepackage{natbib}
\def\bSig\mathbf{\Sigma}

\newcommand{\indep}{\perp \!\!\! \perp}
\usepackage{amssymb}
\usepackage{mathrsfs}
\usepackage{enumerate} % for using bianhao in enumerating
\usepackage{amsmath}
\usepackage{tikz} % for drawing causal graph
\usepackage{setspace}
\usepackage{graphicx}
\usepackage{subcaption}
\usepackage{natbib}
\usepackage{hyperref}
\newtheorem{assumption}{Assumption}
\def\nano{\scriptscriptstyle}
\newcommand\lo[1]{_{\nano #1}}

\def\trans{^{\prime}}
\newcommand\spn[1]{{\mathcal S} ({#1})}
\newcommand\cs[1]{{\mathcal S}\lo {#1|X}}
\newcommand\udex[1]{^{\raisebox{1.2pt}{\mbox{$\nano #1$}}}}

\def\inv{\udex {-1}}
\def\msir{\M\lo {\mathrm{SIR}}}
\def\msave{\M\lo {\mathrm{SAVE}}}
\newcommand\real{{\mathbb R}}

\def\bop{O\lo P}
\def\sop{o\lo P}
\def\var{{\mathrm {Var}}}
\def\cov{{\mathrm {Cov}}}

\def\As{{\mathcal {A}}}
\def\M{{\mathrm {M}}}

\def\Fs{{\mathcal F}}
\def\As{{\mathcal {A}}}
\def\trans{^{\prime}}
\def\nano{\scriptscriptstyle}
\renewcommand\cs[1]{{\mathcal S}\lo {#1|X}}
\def\Fs{{\mathcal F}}
\def\M{{\mathrm {M}}}
\def\bop{O\lo P}
\def\sop{o\lo P}
\def\inv{\udex {-1}}

\def\Ms{{\mathcal {M}}}
\def\Ns{{\mathcal {N}}}
\def\comp{^{\mbox{\tiny{\sf C}}}}
\usetikzlibrary{shapes,decorations,arrows,calc,arrows.meta,fit,positioning}
\tikzset{
	-Latex,auto,node distance =1 cm and 1 cm,semithick,
	state/.style ={ellipse, draw, minimum width = 0.7 cm},
	point/.style = {circle, draw, inner sep=0.04cm,fill,node contents={}},
	bidirected/.style={Latex-Latex,dashed},
	el/.style = {inner sep=2pt, align=left, sloped}
}
\usepackage[plain,noend]{algorithm2e}
\usepackage{lastpage}
\firstpageno{1}

\begin{document}

\title{An exhaustive selection of sufficient adjustment sets for causal inference}

\author{\name Wei Luo \email luoluowei@gmail.com \\
       \addr Center for Data Science\\
       Zhejiang University \\
       Hangzhou, Zhejiang, China
       \AND
       \name Fei Qin \email feiqin95@gmail.com \\
       \addr Department of Mathematics\\
       Hong Kong Baptist University\\
       Kowloon, Hong Kong, China
       \AND
       \name Lixing Zhu \email lzhu@bnu.edu.cn \\
       \addr Center for Statistics and Data Science\\
       Beijing Normal University at
Zhuhai\\
       Zhuhai, Guangdong, China}

%\editor{}

\maketitle

\begin{abstract}  
A subvector of predictor that satisfies the ignorability assumption, whose index set is called a sufficient adjustment set, is crucial for conducting reliable causal inference based on observational data. In this paper, we propose a general family of methods to detect all such sets for the first time in the literature, with no parametric assumptions on the outcome models and with flexible parametric and semiparametric assumptions on the predictor within the treatment groups; the latter induces desired sample-level accuracy. We show that the collection of sufficient adjustment sets can uniquely facilitate multiple types of studies in causal inference, including sharpening the estimation of average causal effect and recovering fundamental connections between the outcome and the treatment hidden in the dependence structure of the predictor. These findings are illustrated by simulation studies and a real data example at the end. 
%\blindtext
\end{abstract}


\section{Introduction} \label{sec: intro}

Causal inference commonly refers to estimating the causal effect of a treatment variable on an outcome variable of interest from observational studies. Usually, the treatment variable $T$ is binary, under which the potential outcome framework \citep{rubin1974} is a popular tool in the literature to formulate the causal inference theory. The potential outcomes, denoted by $Y(1)$ and $Y(0)$, are the outcome variables if a subject hypothetically receives the treatment (with $T=1$) and does not receive the treatment ($T=0$), respectively. The average causal effect is then defined as $E\{Y(1) - Y(0)\}$, and, given a predictor $X \in \real\udex p$ that measures the subject's characteristics, the conditional average causal effect is defined as $E\{Y(1) - Y(0) | X\}$. Because each individual falls into exactly one treatment group in reality, i.e. indexed by $T$, both $Y(1)$ and $Y(0)$ are observable only in the corresponding treatment group. To address their missing, two assumptions are commonly adopted in the literature: first, the ignorability assumption
\begin{equation} \label{as: ignorability}
Y(t) \indep T \mid X \quad \mbox{ for } t=0,1,
\end{equation}
which states that the treatment is randomly assigned regardless of $Y(0)$ or $Y(1)$ given $X$ \citep{rosenbaum1983}, second, the common support condition
\begin{align}\label{as: common supp}
\Omega(X|T=1) = \Omega(X|T=0),
\end{align}
which avoids extrapolation when using each treatment group to model the entire population \citep{rosenbaum1983}. Here, $\Omega(\cdot)$ denotes the support of a distribution. Under (\ref{as: ignorability}) and (\ref{as: common supp}), $E\{Y(t)|X\}$ is identical to $E\{Y(t)|X, T=t\}$ for each $t=0, 1$. Since the latter are estimable by the observed sample, as are the aforementioned causal effects.

To secure the ignorability assumption (\ref{as: ignorability}), researchers often collect a large-dimensional $X$ that includes all the confounders affecting both $Y(t)$ and $T$. This however hinders interpretation and more importantly triggers the ``curse-of-dimensionality" in fitting $E\{Y(t)|X, T=t\}$, generating unreliable causal effect estimation. Consequently, causal inference can be facilitated if one finds a subvector of $X$, denoted by $X\lo A$ where $A \subset \{1,\ldots, p\}$, that satisfies
\begin{equation} \label{as: vs ytx}
Y(t) \indep T \mid X\lo A
\end{equation}
for $t=0$ or $1$. Suppose ${A\lo 0}$ satisfies (\ref{as: vs ytx}) for $Y(0)$ and ${A\lo 1}$ satisfies (\ref{as: vs ytx}) for $Y(1)$. The average causal effect $E\{Y(1)-Y(0)\}$ can then be rewritten as
\begin{align}\label{eq: ate}
E[E\{Y(1)|X\lo {A\lo 1}, T= 1\} - E\{Y(0) | X\lo {A\lo 0},T=0\}],
\end{align} 
and can be estimated accurately if both $A\lo 0$ and $A\lo 1$ have small cardinalities. Following \cite{guo2022}, we call any $A$ that satisfies (\ref{as: vs ytx}) for $t=0$ or $1$ a sufficient adjustment set. Denote the collection of such sets for the corresponding $t$ by $\As\lo t$. Due to the potential complexity of causal effect, $\As\lo 1$ can differ freely from $\As\lo 0$, so it is suitable to discuss them separately. Hereafter, we refer to $t$ as each of $0$ and $1$ if no ambiguity is caused. 

\begin{figure}[htbp]
	\centering
		\centerline{
			
			\begin{tikzpicture}
				\node (1) at (-1.5,1) {$X\lo 1$};
				\node (3) at (0,1) {$X\lo 3$};
				\node (2) at (1.5,1) {$X\lo 2$};
				\node (4) at (-1.5,0) {$X\lo 4$};
				\node (5) at (0,0) {$X\lo 5$};
				\node (6) at (1.5,0) {$X\lo 6$};
				\node (7) at (-3,1) {$Y(t)$};
				\node (8) at (3,1) {$T$};
    
				\path (1) edge (7);
				\path (4) edge (7);
				\path (3) edge (1);
				\path (3) edge (2);
				\path (4) edge (5);
				\path (6) edge (5);
				\path (2) edge (8);
				\path (6) edge (8);
			\end{tikzpicture}}
		%\caption{\label{fig3a}(a)}

	\centering
\caption{An example of the directed acyclic graph of $(Y(t), T, X)$, where $X= (X\lo 1,\ldots, X\lo p)\trans$.}
\label{fig: intro}
\end{figure}
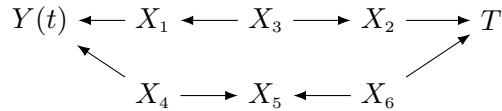

In the literature, special elements of $\As\lo t$ have been widely studied rooted in the sparsity assumption on the treatment model $T|X$ and/or the outcome model $Y(t)|X$ for $t=0, 1$. For example, \cite{weitzen2004} assumed sparsity on $T|X$, that is,
\begin{equation} \label{as: vs tx}
T \indep X \mid X\lo B,
\end{equation}
under which $B$ satisfies (\ref{as: vs ytx}) and thus is a sufficient adjustment set. Similarly, \cite{shortreed2017} and \cite{tang2023} assumed the sparsity of $Y(t)|X$ in terms of
\begin{equation} \label{as: vs yx}
Y(t) \indep X \mid X\lo C,
\end{equation}
where $C$  is also a sufficient adjustment set satisfying (\ref{as: vs ytx}). Under both (\ref{as: vs tx}) and (\ref{as: vs yx}), \cite{vanderweele2011} studied another sufficient adjustment set $B \cup C$, and \cite{de2011} considered two special locally minimal sufficient adjustment sets starting with variable selection on (\ref{as: vs tx}) and (\ref{as: vs yx}), respectively, where the local minimality of $A$ means that no proper subset of $A$ satisfies (\ref{as: vs ytx}). There are generally non-unique and even numerous locally minimal sets in $\As\lo t$; see the directed acyclic graph in Figure \ref{fig: intro} where $\{1\}$, $\{2\}$, and $\{3\}$ are such sets with only the first two detectable by \citeauthor{de2011}'s approach. We refer to \cite{guo2022} for other related works that rely on prior knowledge about the causal graph. 

As pointed out in \cite{witte2019} and \cite{henckel2022}, the optimal sufficient adjustment set varies with the specific causal effect of interest. For example, the formula (\ref{eq: ate}) suggests that the locally minimal sets are best among all in $\As\lo t$ to estimate the average causal effect. A thorough investigation to this end however requires estimating all the locally minimal sets in $\As\lo t$, which, as seen above, remains unavailable in the literature. Besides causal effect estimation, often the purpose of causal inference is to understand the fundamental connection between $Y(t)$ and $T$ revealed by the subject's personal characteristics $X$, which hinges on the dependence within $X$ and cannot be detected by (\ref{as: vs tx}) and (\ref{as: vs yx}) that only model $Y(t)|X$ and $T|X$. These together urge the need of exhaustively recovering $\As\lo 0$ and $\As\lo 1$ in order to facilitate the widest subsequent causal inference, which we will study in this paper for the first time in the literature. The details of how to use $\As\lo 0$ and $\As\lo 1$, particularly in uncovering the deep dependence mechanism between $Y(t)$ and $T$, will be elaborated in \S\ref{sec: use sas} later. %As we are aware of, our work is the first attempt in the literature to address   
%As we know of, this is the first attempt of doing so in the literature. 
%The various types of subsequent causal effect estimation given $\As\lo 1$ and $\As\lo 0$ can be conducted similarly to (\ref{eq: cate}) and are beyond our scope.
%We aim to recover all the qualified $X\lo A$ with no parametric assumptions on $Y(t)|X$ and mild assumptions on $T|X$ in this paper,

\vspace{-0.35cm}

\vspace{-0.2cm}

Because  (\ref{as: vs ytx}) involves two responses and we only focus on their connection via $X$, it differs from and is more complex than the conventional variable selection that only involves one response, e.g. (\ref{as: vs yx}). First, it does not require the sparsity assumption (\ref{as: vs tx}) or (\ref{as: vs yx}). For example, suppose $X$ affects $Y(t)$ and $T$ through some non-sparse $\beta\lo {Y(t)}\trans X$ and $\beta\lo T\trans X$, respectively, and $\beta\lo {Y(t)}\trans X \indep \beta\lo T\trans X$; then both (\ref{as: vs tx}) and (\ref{as: vs yx}) fail but $Y(t)$ is marginally independent of $T$. For the most generality of the proposed work, we do not assume (\ref{as: vs tx}) or (\ref{as: vs yx}) (except in \S $5$), despite that they are common adopted in the literature as mentioned above. Second, while any superset of $C$ in (\ref{as: vs yx}) also satisfies (\ref{as: vs yx}), this can fail for (\ref{as: vs ytx}); see Figure \ref{fig: intro} again where $\{1\}$ satisfies (\ref{as: vs ytx}) but $\{1, 5\}$ does not. Similarly, while any $A$ and $B$ that satisfy (\ref{as: vs yx}) implies that $A \cap B$ also satisfies (\ref{as: vs yx}), this nesting property generally fails for (\ref{as: vs ytx}), leading to the aforementioned non-uniqueness of the locally minimal sets in $\As\lo t$. Therefore, to recover $\As\lo 1$ and $\As\lo 0$ in general, one must evaluate the validity of (\ref{as: vs ytx}) exhaustively for all $A \subset \{1,\ldots, p\}$, and any smoothing procedure that mimics the lasso method \citep{tibs1996} can be inefficient due to the presence of multiple minima. Simplification in some special cases will be discussed in \S\ref{sec: use sas}.

The evaluation of (\ref{as: vs ytx}) requires a quantitative criterion that has negligible values if and only if $A \in \As\lo t$. Because $X\lo A$ can be large-dimensional when $A$ is close to $\{1,\ldots, p\}$, certain modeling assumptions are necessary to avoid the ``curse-of-dimensionality" for the corresponding criteria. To this end, we modify the dual inverse regression methods recently proposed by \cite{luo2022}, which conducts dimension reduction concerning (\ref{as: vs ytx}) with $Y(t)$ replaced by a general response and with $X\lo A$ replaced by a free low-dimensional linear combination $\beta\trans X$. To address the missing of $Y(t)$ and the dimensionality of $X\lo A$, we start with assuming normality on $X|T$ with possibly heterogeneous covariance matrix, and then generalize it to a semiparametric Gaussian copula that allows skewness and heavy tails, etc. The proposed criteria are always model-free on $Y(t)|X$ and meanwhile accurate in the sample level, given which we use the ridge ratio method, (see, e.g., a relevant reference \cite{xia2015}), to detect $\As\lo t$ as an illustration among many other choices.
%This method only requires $X$ to have a multivariate normal distribution, and is otherwise model-free. However, it is targeted at a general response $Y$ rather than the potential outcome $Y(t)$ in causal inference, and therefore did not carefully address the missing of these outcomes. By contrast, we will incorporate and discuss thoroughly the corresponding assumptions and prices, particularly .etc.
%In case the data are collected from experimental studies where $Y(t)$ is completely observed, we can again make $X|T$ model-free and instead assume the marginal $X$ to be normal or following a Gaussian copula.
%Regarding variable selection as dimension reduction with specific reduced sps, the proposed work is also related to the a recent work \cite{luo2021} 
%whose tail tends to identify $\As\lo t$, 

To summarize, the proposed methods innovatively recover the collections of all the sufficient adjustment sets without assuming sparsity on $Y(t)|X$ or $T|X$. They are always model-free on $Y(t)|X$, and meanwhile accurate and easily implementable by adopting flexible parametric or semiparametric model on $X|T$. The collections of the sufficient adjustment sets facilitate multiple types of causal inference, including sharpening the causal effect estimation and discovering fundamental dependence mechanism between $Y(t)$ and $T$. In the following, we will first focus on the estimation of $\As\lo t$'s and then discuss how to use them for subsequent causal inference. The theoretical proofs are deferred to the Appendix. For simplicity, we assume $X$ to be numerical with zero mean throughout the article.

\section{A brief review of dual inverse regression} \label{sec: review}

\newcommand\bblue[1]{\textcolor{blue}{{\huge {#1}}}}

We first review the dual inverse regression methods \citep{luo2022} that inspire the proposed method. Let $W$ be a general, presumably fully observed, response in place of $Y(t)$. \cite{luo2022} assumed and estimated a low-dimensional linear combination $\beta\trans X$ that satisfies
\begin{equation} \label{as: dim reduction ytx}
W \indep T \mid \beta \trans X.
\end{equation} 
This coincides with  (\ref{as: vs ytx}) if $W$ is replaced with $Y(t)$ and $\beta\trans X$ is restricted to be some $X\lo A$. To tackle (\ref{as: dim reduction ytx}), \cite{luo2022} employed the candidate matrices in the inverse regression methods \citep{li1991, cook1991, li2007} as the basic elements, which were originally designed for the conventional sufficient dimension reduction with respect to an individual response. Therefore, to explain the dual inverse regression methods, we first briefly review sufficient dimension reduction and the associated inverse regression methods. As seen later, this is also the key to the proposed methods for relaxing the sparsity assumptions (\ref{as: vs tx}) and (\ref{as: vs yx}).

%Because the validity of (\ref{eq: dim reduction interact}) is invariant of invertible column transformations of $\beta$, \cite{luo2021} regarded the column space of $\beta$, denoted by $\spn \beta$, as the parameter of interest for this type of dimension reduction problem and call it a locally efficient dimension reduction subspace

Using $W$ as the response, sufficient dimension reduction assumes the existence of a low-dimensional $\beta\lo W\trans X$ such that
\begin{align}\label{as: dim reduction yx}
W \indep X \mid \beta\lo W\trans X.
\end{align}
For identifiability, \cite{cook1998} introduced the central subspace $\cs W$ as the parameter of interest, defined as the unique subspace of $\real\udex p$ with minimal dimension whose arbitrary basis matrix satisfies (\ref{as: dim reduction yx}). To estimate $\cs W$, the aforementioned inverse regression methods commonly use the moments of $X|W$ to construct a matrix-valued parameter, called the candidate matrix and denoted by $\M$, whose column space $\spn \M$ coincides with $\cs W$ under mild conditions. For example, the candidate matrices for the sliced inverse regression \citep{li1991} and the sliced average variance estimator \citep{cook1991} are, respectively, 
\begin{align*}
\msir = \Sigma \inv \{E(X\mid W\lo S = h)\}\lo {h=1,\ldots, H}, \quad \msave = \Sigma\inv \{\Sigma - \var (X\mid W\lo S = h)\}\lo {h=1,\ldots, H},
\end{align*}
where $\Sigma$ denotes the covariance matrix of $X$, $(v\lo h)\lo {h=1,\ldots, H}$ denotes $(v\lo 1,\ldots, v\lo H)$ for any matrices $v\lo 1,\ldots, v\lo H$, and $W\lo S$ coincides with $W$ if $W$ is discrete and otherwise slices $W$ as $W\lo S = h$ whenever $W$ falls between its $(h-1)/H$th and $h/H$th sample quantiles, $H$ being prefixed. Using $W\lo S$ induces a $n\udex {1/2}$-consistent estimator $\widehat \M$ based on the marginal and slice sample moments of $X$ \citep{li1991, cook1991}, and its theoretical guarantee is that $W|X$ and $W\lo S|X$ share the same central subspace as long as the slicing is fine enough \citep{zhu1995asymptotics, li2007asymptotics,li2018}.

To ensure $\spn \M = \cs W$, the sliced inverse regression requires the linearity condition on $X$ or $X|W$, that is, with $\mu\udex {(W)}$ and $\Sigma\udex {(W)}$ being the mean and the covariance matrix of $X|W$, and $P(\Sigma, \beta\lo W) \equiv \beta\lo W(\beta\lo W\trans \Sigma \beta\lo W)\inv \beta\lo W\trans \Sigma$ and $P(\Sigma\udex {(W)}, \beta\lo W)$ likewise being the projection matrices,
\begin{align}\label{as: linearity}
E(X | \beta\lo W\trans X) = P\trans(\Sigma, \beta\lo W) X \mbox{ or } E(X | \beta\lo W\trans X, W) = P\trans(\Sigma\udex {(W)}, \beta\lo W) (X - \mu\udex {(W)}) + \mu\udex {(W)}.
\end{align}
Besides (\ref{as: linearity}), the sliced average variance estimator and other inverse regression methods require the constant variance condition on either $X$ or $X|W$,  as formulated by
\begin{align}\label{as: const var}
\var(X | \beta\lo W\trans X) = \Sigma - \Sigma P(\Sigma, \beta\lo W) \mbox{ or }
\var(X | \beta\lo W\trans X, W) = \Sigma\udex {(W)} - \Sigma\udex {(W)} P(\Sigma\udex {(W)}, \beta\lo W).
\end{align} 
When $p$ is large and $\cs W$ is low-dimensional, both (\ref{as: linearity}) and (\ref{as: const var}) hold approximately \citep{diaconis1984, hall1993}. Otherwise, they require $X$ or $X|W$ to have a multivariate normal distribution, considering the freedom of unknown $\cs W$. Depending on whether symmetric effect may exist in $W|X$, etc., researchers can choose the working inverse regression methods in a data-driven manner; see \cite{li2018} for details. For simplicity, throughout the article we assume that the working $\M$ always spans the central subspace, subject to (\ref{as: linearity}) and (\ref{as: const var}).
% to have an asymmetric effect on $W$. 
%the working inverse regression method has been appropriately chosen so that the corresponding
%low-dimensional linear combination, which restricts the applicability of the inverse regression met hods.
% is the projection matrix onto $\cs W$ under the inner product $\langle u, v\rangle\lo \Sigma = u\trans \Sigma v$
% $W$ to have a large enough number of categories if it is discrete or categorical
%In addition, it requires asymmetry of $\beta\lo W\trans X | W$. 
%under which they can detect symmetric or other subtle effects of $\beta\lo W\trans X$ on $W$. 

Let $\M\lo {W}$ and $\M\lo T$ be the candidate matrices applied to $W|X$ and $T|X$, respectively, induced from potentially different inverse regression methods. \cite{luo2022} chracterized (\ref{as: dim reduction ytx}) by
\begin{align} \label{eq: dir}
\M\lo {W}\trans \Sigma P(\Sigma, \beta) \M\lo T = \M\lo {W}\trans \Sigma \M\lo T.
\end{align}
A dual inverse regression method then solves (\ref{eq: dir}) with $\Sigma$, $\M\lo {W}$, and $\M\lo T$ replaced by their estimators. To ensure the coincidence between the solutions to (\ref{eq: dir}) and the reduced predictors for (\ref{as: dim reduction ytx}), \cite{luo2022} adopted a regularity assumption that equates (\ref{as: dim reduction ytx}) with
\begin{align}\label{as: dim reduction by bt bx}
\beta\lo W\trans X \indep \beta\lo T\trans X \mid \beta\trans X,
\end{align}
where $\beta\lo T$ spans $\cs T$. Similarly to (\ref{as: linearity}) and (\ref{as: const var}), they also regulated $X$ such that (\ref{as: dim reduction by bt bx}) is equivalent to $\cov (\beta\lo {W}\trans X, \beta\lo T\trans X \mid \beta\trans X) = 0$ for any $\beta$ that solves (\ref{eq: dir}), which again holds approximately if $p$ is large and $(\beta\lo W, \beta\lo T, \beta)\trans X$ is low-dimensional, and otherwise requires $X$ to be normally distributed. We will elaborate variations of these assumptions with the missing of $Y(t)$ carefully addressed and without the dimension reduction assumption (\ref{as: dim reduction ytx}). 

As pointed out in \cite{luo2022}, the dual inverse regression methods do not rely on the sufficient dimension reduction assumption (\ref{as: dim reduction yx}), as a candidate matrix can be implemented even if the corresponding central subspace is $\real\udex p$. However, the existence of proper solutions to (\ref{eq: dir}) requires one of $\M\lo W$ and $\M\lo T$ to have reduced row rank, which means that either $\cs {W}$ or $\cs T$ must be a proper subspace of $\real\udex p$. This implicit assumption is inherited from the aforementioned regularity assumption related to (\ref{as: dim reduction by bt bx}), and it persists in the proposed methods; see detail in \S\ref{sec: method} later.

%\begin{assumption}\label{as: y t bx = byx bt bx}
%For any $\beta \in \real\udex {p \times d}$ with $d< p$, if $\beta\lo {W} \trans  X \not\!\perp\!\!\!\perp \beta\lo T \trans  X \mid \beta \trans  X$, then there exist measurable functions $h({W})$ and $g(T)$ with $E^2 \{h({W})|\beta\lo {W} \trans  X \}< \infty$ and $E^2 \{g(T)|\beta\lo T \trans  X \}< \infty$ such that $cov[E\{h({W})|\beta\lo {W} \trans  X\}, E\{g(T)|\beta\lo T \trans  X\} |\beta\trans X]$ is not degenerate at zero.
%\end{assumption}
%\medskip
%However, this assumption is unnecessary for (\ref{as: dim reduction ytx}).
%The dual inverse regression method then solves (\ref{eq: dir}) with $\M\lo Y$ and $\M\lo T$ replaced by their sample estimators.
%This assumption fails in some extreme cases, e.g. if $X\lo 1$ has a symmetric distribution with respect to zero and $Y = X\lo 1$ and $T = {\mathrm {sign}}(X\lo 1)$, and it is satisfied in general.
%it is again approximately satisfied if $p$ is large and $\spn {\beta\lo Y, \beta\lo T, \beta}$ is low-dimensional, but otherwise it

\section{The proposed method} \label{sec: method}

We first introduce some notations. Let $\Fs$ be the collection of all the subsets of $\{ 1,\ldots,p\}$, which includes both $\As\lo 1$ and $\As\lo 0$ as sub-collections. For any $A \in \Fs$, let $X\lo A$ be the subvector of $X$ indexed by $A$, and let $X\lo {-A}$ be the rest of $X$; let $\Sigma\lo {A, A}$, $\Sigma\lo {-A, -A}$, and $\Sigma\lo {A, -A}$ be the covariance matrix of $X\lo A$, that of $X\lo {-A}$, and that between $X\lo A$ and $X\lo {-A}$, respectively. For a candidate matrix $\M\lo W$, let $\M\lo {W, A}$ be its submatrix consisting of rows indexed by $A$, and let $\M\lo {W,-A}$ be the rest of $\M\lo W$. We use $I\lo A$ to denote the submatrix of $I\lo p$ consisting of its columns indexed by $A$, where $p$ is omitted from the subscript. Denote the empty set by $\emptyset$.    
%Same as above, we refer $Y(t)$ to each of $Y(0)$ and $Y(1)$ if no ambiguity is caused. 

Recall from \S 1 that the key to selecting the sufficient adjustment sets is to construct a good criterion for (\ref{as: vs ytx}). This naturally motivates us to extend the dual inverse regression methods. To this end, we echo \cite{luo2022} to regulate  $Y(t)|\beta\lo {Y(t)}\trans X$ and $T|\beta\lo T\trans X$ as follows. 

\def\nindep{\not\!\perp\!\!\!\perp}

\medskip

\begin{assumption} \label{as: y t xa = byx bt xa} 
If $\beta\lo {Y(t)}\trans X \not\indep \beta\lo T\trans X \mid X\lo A$, then there must exist some $h(\cdot): \real \rightarrow \real$ with finite $E\{ h\udex 2 (Y(t)) | X \}$ such that $\cov[E\{ h(Y(t)) | \beta\lo {Y(t)}\trans X\}, E\{T | \beta\lo T\trans X\} | X\lo A]$ is not degenerate at zero. 
\end{assumption}

\medskip

\noindent
The generality of Assumption \ref{as: y t xa = byx bt xa} resembles the relative discussion in \cite{luo2022}. An exception is that when both $\cs {Y(t)}$ and $\cs T$ are $\real\udex p$, Assumption \ref{as: y t xa = byx bt xa} would imply the failure of (\ref{as: vs ytx}) for any proper subset $A$ of $\{1,\ldots,p\}$, following $X \not\indep X \mid X\lo A$. In this case, the validity of Assumption \ref{as: y t xa = byx bt xa} depends on whether we still assume the existence of a proper sufficient adjustment set. Because any model that satisfies the latter must be subtle in this case, and a proper central subspace is commonly accepted along with the wide applicability of sufficient dimension reduction, which particularly allows non-sparse effects of $X$, we assume that at least one of $\cs {Y(t)}$ and $\cs T$ is a proper subspace of $\real\udex p$. This sufficient dimension reduction assumption is omitted from the rest of the article, as it is implied implicitly by Assumption \ref{as: y t xa = byx bt xa}.

As mentioned in \S 1, we must address the missing of $Y(t)$ indexed by $T=t$, which is briefly mentioned but not carefully dealt with in \cite{luo2022} due to the use of general response $W$. Under the ignorability assumption (\ref{as: ignorability}) and the common support condition (\ref{as: common supp}), $\cs {Y(t)}$ coincides with ${\mathcal S}\lo {Y(t)|(X, T=t)}$ \citep{luo2017}; the latter, as defined on the sub-population $T=t$, is readily estimable. Thus, we instead recover ${\mathcal S}\lo {Y(t)|(X, T=t)}$ by inverse regression, e.g. with 
\begin{align*}
(\Sigma\udex {(t)})\inv [E\{X| Y\lo {S}(t) = h, T=t\}]\lo {h=1,\ldots, H}
\end{align*}
for the sliced inverse regression, where $\Sigma\udex {(t)}$ denotes the covariance matrix of $X|(T=t)$. The modified candidate matrix is still denoted by $\M\lo {Y(t)}$, if no ambiguity is caused. For $\M\lo {Y(t)}$ to span ${\mathcal S}\lo {Y(t)|(X, T=t)}$ and thus also $\cs Y$, and for the same $\M\lo T$ as in \S2 to span $\cs T$, we guarantee the linearity condition (\ref{as: linearity}) and the constant variance condition (\ref{as: const var}) on $X|T$ by adopting 

\vspace{.2cm}

\begin{assumption}\label{as: normal x|t}
For $t= 0, 1$, $X\mid (T=t)$ follows a multivariate normal distribution.
\end{assumption}

\vspace{.2cm}

\noindent
By the Bayes Theorem, Assumption \ref{as: normal x|t} implies a parametric model on $T | X$, which will reduce to the logistic regression if we restrict $\Sigma\udex {(0)} = \Sigma\udex {(1)}$. The reason that we impose strict normality on $X|T$ rather than using the approximation results mentioned in \S 2 is that, first, we allow $\cs {Y(t)}$ and $\cs T$ to be large or even $\real\udex p$ to incorporate complex structures in $Y(t)|X$ and $T|X$ (in terms of free $\Sigma\udex {(t)}$), second, as seen later, a normal $X|T$ induces simple criterion for (\ref{as: vs ytx}), where $X\lo A$ can be large-dimensional. Relaxation of Assumption \ref{as: normal x|t} will be proposed in \S\ref{sec: extension}. 
%Since we require Condition (\ref{as: uncor means indep}) for $\beta$ being a free subset of columns of $I\lo p$, whose number of columns can be close to $p$, the approximation results in \cite{diaconis1984} and \cite{hall1993} are no longer applicable.

Because $X$ is marginally non-normal under Assumption {\ref{as: normal x|t}}, (\ref{eq: dir}) no longer characterizes (\ref{as: vs ytx}) if we simply replace $\beta$ with $I\lo A$. Instead, we characterize (\ref{as: vs ytx}) by 
\begin{align} \label{eq: dir A}
\M\lo {Y(t)}\trans \Sigma\udex {(s)} P(\Sigma\udex {(s)}, I\lo A) \M\lo T =  \M\lo {Y(t)}\trans \Sigma\udex {(s)} \M\lo T \quad \mbox{for both } s= 0, 1.
\end{align}
This is based on an equivalence property that, under the common support condition (\ref{as: common supp}),
\begin{align}\label{as: vs byx btx xa w/o t}
\beta\lo {Y(t)}\trans X \indep \beta\lo T\trans X \mid (X\lo A, T) \,\, \Leftrightarrow \,\, \beta\lo {Y(t)}\trans X \indep \beta\lo T\trans X \mid X\lo A \quad \mbox{for any } A\in \Fs.
\end{align}
Namely, the former in (\ref{as: vs byx btx xa w/o t}) is on the sub-population indexed by $T$ and is what (\ref{eq: dir A}) actually characterizes under Assumption \ref{as: normal x|t}, as (\ref{eq: dir}) characterizes (\ref{as: dim reduction by bt bx}) under normal $X$; the latter is on the entire population and coincides with (\ref{as: vs ytx}) under the regularity conditions, as (\ref{as: dim reduction by bt bx}) coincides with (\ref{as: dim reduction ytx}). The key to (\ref{as: vs byx btx xa w/o t}) is the definition of $\cs T$ spanned by $\beta\lo T$; see detail in the Appendix.
%the varying $s$ represents the randomness of $T$ in the first conditional independence in (\ref{as: vs byx btx xa w/o t}).
%which is justified in the following under all the regularity assumptions above. 
%where $P(\Sigma\udex {(s)}, I\lo A)$ again denotes the projection matrix $I\lo A (\Sigma\udex {(s)}\lo {A,A})\inv I\lo A\trans \Sigma\udex {(s)}$. 

\medskip

\begin{theorem} \label{thm: normal x|t}
Under the ignorability assumption (\ref{as: ignorability}), the common support condition (\ref{as: common supp}), and Assumptions \ref{as: y t xa = byx bt xa}-\ref{as: normal x|t}, for any $A \in \Fs$, (\ref{eq: dir A}) holds if and only if (\ref{as: vs ytx}) holds. that is, $A \in \As \lo t$.
\end{theorem}

\medskip

By Theorem \ref{thm: normal x|t}, (\ref{eq: dir A}) can induce good criteria for detecting the sufficient adjustment sets $\As\lo t$ subject to the normality of $X|T$. As an illustration, we introduce $f\lo t: \Fs \rightarrow [0, \infty)$ with, if $A$ is nonempty, 
\begin{align}\label{eq: f normal x|t}
f\lo t (A) = \sum\lo {s=0,1}\| \M\lo {Y(t), -A}\trans \{\Sigma\udex {(s)}\lo {-A, -A} - \Sigma\udex {(s)}\lo {-A, A} (\Sigma\udex {(s)}\lo {A, A})\inv  \Sigma\udex {(s)}\lo {A, -A}\} \M\lo {T, -A} \|
\end{align}
 and with $f\lo t(\emptyset) = \sum\lo {s=0,1} \|M\lo {Y(t)}\trans \Sigma\udex {(s)} M\lo {T}\|$, where $\|\cdot\|$ denotes the spectral norm. The solution set of $f\lo t(A) = 0$, which also minimizes $f\lo t(\cdot)$, identifies $\As\lo t$. For $f\lo t (A) = 0$, either $\M\lo {Y(t), -A}$ or $\M\lo {T,-A}$ must have reduced row rank, which complies with the assumption above that either $\cs {Y(t)}$ or $\cs T$ is a proper subspace of $\real\udex p$. Nonetheless, these matrices can still be non-sparse, indicating the non-sparsity of $Y(t)|X$ and $T|X$ mentioned in \S 1. Again, this generality is not permitted in the existing approaches, even though they only detect special members of $\As\lo t$ \citep{shortreed2017, tang2023}.

\begin{figure}[htbp]
\centerline{\includegraphics[width=0.4\textwidth]{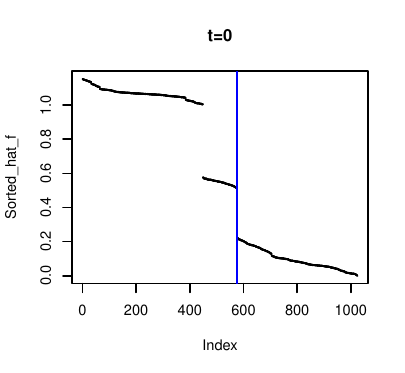}}
\caption{The plot of $\widehat f\lo 0(A\lo k\udex {(0)})$'s for Model 2 in \S\ref{sec: sim}, where $(n,p)=(400,10)$ and $\As\lo 0$ has $448$ sets.} 
\label{fig: scree plot}
\end{figure}

Let $\widehat \Sigma\udex {(s)}$ be the sample covariance matrix of $X|(T=s)$ for $s=0, 1$, and let $\widehat \M\lo {Y(t)}$ and $\widehat \M\lo T$ be the consistent estimators of $\M\lo {Y(t)}$ and $\M\lo T$ reviewed in \S\ref{sec: review}. We approximate $f\lo t(\cdot)$ by plugging these matrix estimates into (\ref{eq: f normal x|t}). Based on the resulting $\widehat f\lo t(\cdot)$, we choose the ridge ratio method \citep{xia2015} to detect $\As\lo t$ for simplicity. Let $\{A\lo k\udex {(t)}: k =1,\ldots, 2\udex p\}$ be an rearrangement of $\Fs$ such that $\widehat f\lo t(A\lo {k}\udex {(t)}) \geq \widehat f\lo t(A\lo {k+1}\udex {(t)})$ for each $k$. For large samples, $\widehat f\lo t(A\lo {k}\udex {(t)})$'s convey a scree plot with $\As\lo t$ mostly indexing its tail; see Figure \ref{fig: scree plot} generated from Model 2 in \S\ref{sec: sim} with $(n,p)=(400,10)$ and $\As\lo 0$ consisting of $448$ sets. Define the ridge ratio $R\lo t(\cdot): \{0,\ldots, 2\udex p-1\} \rightarrow \real$ as 
\begin{equation} \label{eq: taut}
R\lo t (k) = \{\widehat{f}\lo t (A\lo {k+1}\udex {(t)}) +c\lo {n}\} \, / \, \{\widehat{f}\lo t (A\lo {k}\udex {(t)}) + c\lo n\} \quad \mbox{for } k>0,
\end{equation}
where $c\lo n > 0$, and $R\lo t (0) = c\lo 0\in (0, 1)$. The use of $R\lo t(0)$ is to incorporate the extreme case that $f\lo t(\cdot)$ is constantly zero, which occurs if (\ref{as: vs ytx}) holds for all $A \in \Fs$. We use the minimizer of $R\lo t(\cdot)$, denoted by $\tau(t)$, to distinguish the tail of the scree plot above, namely
\begin{align}\label{eq: hat At}
\widehat \As\lo t = \{A\lo k\udex {(t)}: k > \tau(t)\}.
\end{align}
The consistency of $\widehat \As\lo t$ follows from that of $\widehat \M\lo {Y(t)}$, $\widehat \M\lo T$, and $\widehat \Sigma\udex {(s)}$, subject to appropriate $c\lo n$. 
%where the regularity term $c\lo n$ vanishes in certain rates. We use $\tau(t)$ as the cutoff for the tail of the scree plot, i.e. the set of insignificant values of $\widehat f\lo t (\cdot)$, and estimate $\As\lo t$ by
%%sequence $\tau(t)$ be the minimizer of ridge ratio of the rearranged $\widehat f\lo t(\cdot)$, that is,

\vspace{.2cm}

\begin{theorem} \label{thm: hf=f normal x|t}
Under the ignorability assumption (\ref{as: ignorability}), the common support condition (\ref{as: common supp}), and Assumptions \ref{as: y t xa = byx bt xa} - \ref{as: normal x|t}, we set $c\lo n$ in (\ref{eq: taut}) such that $\max\{n\udex {-1/2} / c\lo n, c\lo n\}= o(1)$, then $\widehat \As\lo t$ coincides with $\As\lo t$ with probability converging to one. %with probability converging to one.
\end{theorem}

\vspace{.2cm}

In the numerical studies later, we set $c\lo 0 = .6$ and $c\lo n = .2 n\udex{-1/2} \log n$ uniformly. When $p$ is large, we instead suggest using the sample splitting technique \citep{zou2020} to address the growing accumulative noise in $\widehat f\lo t(\cdot)$, details omitted. The asymptotic study under the high-dimensional settings is deferred to the Appendix.
%To formulate the consistency of $\widehat \As\lo t$ under these high-dimensional settings, we must borrow from the literature \citep{lin2018, wainwright2019} the asymptotic consistency of $\widehat \M\lo {Y(t)}$, $\widehat \M\lo T$, and $\widehat \Sigma\udex {(s)}$, which regulates the diverging rate of $p$ that varies with the specific inverse regression methods, and adjust the vanishing rate of $c\lo n$ accordingly. The research interest under these settings may also switch to the selection of all the small sufficient adjustment sets instead of the entire $\As\lo t$. 

One issue in implementing $\widehat \As\lo t$ is its computational cost for the exhaustive search over $\Fs$, which grows in an exponential rate of $p$. For example, with $n=400$ and the sliced inverse regression used for both $\M\lo {Y(t)}$ and $\M\lo T$, it takes $\approx .2$ second on a standard desktop when $p$ is $10$ but elevates to $\approx 10$ minutes when $p$ grows to $20$. As mentioned in \S 1, this is generally inevitable due to the complexity of $\As\lo t$. Simplification in certain cases will be discussed in \S\ref{sec: use sas}.

\section{Extension to non-normal {\small {$X\mid T$}}} \label{sec: extension}

We now relax Assumption \ref{as: normal x|t} to allow more general semiparametric models for $X|T$. The key to our extension is the invariance of (\ref{as: vs ytx}) under any component-wise monotone transformation $(\zeta\lo 1 (X\lo 1), \ldots, \zeta\lo p (X\lo p))\trans$, denoted by $\zeta(X)$; that is, for any $A \in \Fs$, (\ref{as: vs ytx}) holds if and only if
\begin{align*}
Y(t) \indep T \mid (\zeta (X))\lo A.
\end{align*}
Consequently, we allow $X$ to have free component-wise distributions and only require its conditional normality on $T$ subject to the corresponding component-wise transformations. This equivalently means a homogeneous semiparametric Gaussian copula model for $X|T$, that is, 
%Again, due to the missing of $Y(t)$ indicated by $T$, a nonparametric $X|T$ cannot induce a good criterion for (\ref{as: vs ytx}) that is meanwhile model-free on $Y(t)|X$. 

\vspace{.2cm}

\begin{assumption}\label{as: gcopula x|t}
There exist (completely unknown) monotone mappings $\zeta\lo 1 (\cdot),\ldots, \zeta\lo p(\cdot)$ such that $\zeta(X)|(T=s)$ has a multivariate normal distribution for both $s=0, 1$.
\end{assumption}
%from $\real$ to $\real$ 

\vspace{.2cm}

\noindent
When $\zeta(X) = X$, this reduces to the normality of $X|T$. The applicability of Gaussian copula in fitting skewed or heavily-tailed distributions has been widely recognized \citep{liu2012}.

Let $\M\lo {Y(t)|\zeta(X)}$ and $\M\lo {T|\zeta(X)}$ be the same matrices as before with $X$ replaced by $\zeta(X)$, and let $\Omega\udex {(s)}$ be the covariance matrix of $\zeta(X)|(T=s)$. We generalize $f\lo t(\cdot)$ in (\ref{eq: f normal x|t}) to 
\begin{align} \label{eq: f g-coupla x|t}
f\lo t\udex *(A) = \sum\lo {s=0,1} \| \M\trans \lo {Y(t) | \zeta(X),-A} \{\Omega\udex{(s)}\lo {-A,-A}-\Omega\udex{(s)}\lo{-A,A}   (\Omega\udex{(s)}\lo{A ,A})\udex{-1}  \Omega\udex{(s)}\lo{A,-A} \} \M\lo {T|\zeta(X),-A} \|
\end{align}
if $A$ is nonempty, and to $f\lo t\udex *(\emptyset) = \sum\lo {s=0,1} \|\M\lo {Y(t)|\zeta(X)}\trans \Omega \udex {(s)} \M\lo {T|\zeta(X)}\|$. Again, we adopt the ignorability assumption (\ref{as: ignorability}), the common support condition (\ref{as: common supp}), and Assumption \ref{as: y t xa = byx bt xa} on $(Y(t), T, \zeta(X))$. The former two are invariant under the transformation $\zeta(\cdot)$; the latter requires ${\mathcal S} \lo {Y(t)|\zeta (X)}$ or ${\mathcal S}\lo {T|\zeta(X)}$ to be a proper subspace of $\real\udex p$, which we assume implicitly. The identification of $\As\lo t$ as the solution set of $f\lo t\udex * (A) = 0$ follows similarly to Theorem \ref{thm: normal x|t}.

%\vspace{.2cm}

%\begin{corollary} \label{cor: g-copula x|t}
%Suppose $X|T$ follows a homogeneous semiparametric Gassian copula model. Under the ignorability assumption (\ref{as: ignorability}), the common support condition (\ref{as: common supp}), and Assumption \ref{as: y t xa = byx bt xa} for $(Y(t), T, \zeta(X))$, $\As\lo t$ coincides with the solution set of $f\lo t(A) = 0$. 
%\end{corollary}  

%\vspace{.2cm}
%\textcolor{red} {QinFei: DO NOT make ridiculous mistakes again!}

To estimate $f\lo t \udex * (\cdot)$, we approximate each $\zeta\lo i(\cdot)$ in two steps. First, within each treatment group $T=s$, we follow the literature \citep{klaassen1997} to use the normal-score estimator to approximate $\zeta\lo i(\cdot)$ up to a linear transformation specified for $s$; that is, let $\widehat \zeta\lo {i, s}(\cdot) = \Phi\inv \{\widehat {F}\udex *\lo {i, s}(\cdot)\}$, where $\Phi (\cdot)$ is the cumulative distribution function of the standard normal distribution, and $\widehat F\udex *\lo {i, s}(\cdot)$ is the empirical cumulative distribution function for $X\lo i$ given $T=s$ multiplied by $(1 + n\lo s\inv)$, $n\lo s$ being the number of observations with $T=s$. Second, as $\widehat \zeta\lo {i,0}(\cdot)$ and $\widehat \zeta\lo {i,1}(\cdot)$ differ by a linear transformation asymptotically, we pool them by transforming $\widehat \zeta\lo {i,1}(\cdot)$ to $\widehat a\lo i \widehat \zeta\lo {i,1}(\cdot) + \widehat b\lo i$, where $(\widehat a\lo i, \widehat b\lo i)$ minimizes the truncated squared loss
\begin{align}\label{eq: gcopula pool}
E\lo n \left[ [\widehat \zeta\lo {i,0} (X\lo i) - \{a\lo i \widehat \zeta\lo {i, 1} (X\lo i) + b\lo i\}]\udex 2 \cdot \delta [\max\{|\widehat \zeta\lo {i,0} (X\lo i)|, |\widehat \zeta\lo {i,1} (X\lo i)|\} < \Phi\inv(.975)] \right]
\end{align}
over $(a\lo i, b\lo i) \in \real\udex 2$, $E\lo n(\cdot)$ being the sample mean and $\delta(\cdot)$ being the indicator function. Under the common support condition (\ref{as: common supp}), the truncation in (\ref{eq: gcopula pool}) still preserves a certain proportion of the observed sample in the estimation. We take $(1-T) \widehat \zeta\lo {i,0}(\cdot) + T \{\widehat a\lo i \widehat \zeta\lo {i,1}(\cdot) + \widehat b\lo i\}$ as the pooled estimate of $\zeta\lo i(\cdot)$, which permits using the same estimators as before for $\M \lo {Y(t) | \zeta(X)}$, $\M\lo {T|\zeta(X)}$, and $\Omega\udex{(s)}$. The resulting $\widehat f\lo t\udex * (\cdot)$ again delivers a consistent $\widehat \As\lo t$ by (\ref{eq: taut}) and (\ref{eq: hat At}). 
%To estimate $\Omega\udex{(s)}$, we still use the commonly adopted Kendall rank correlation coefficients \citep{liu2012}, details omitted.
%Because $\widehat \zeta\lo {i,0}(\cdot)$ is intuitively more accurate than $\widehat \zeta\lo {i,1}(\cdot)$ when $T=0$ and vice versa,

\vspace{.2cm}

\begin{theorem}\label{thm: hf=f gcopula x|t} 
Under the ignorability assumption (\ref{as: ignorability}), the common support condition (\ref{as: common supp}), Assumption \ref{as: gcopula x|t}, and the adjusted Assumption \ref{as: y t xa = byx bt xa} for $\zeta(X)$ specified in Assumption \ref{as: gcopula x|t}, we have $\widehat \As\lo t =\As\lo t$ with probability converging to one if $c\lo {n}$ in (\ref{eq: taut}) satisfies $\max\{ c\lo n, n\udex {-1/2} / c\lo n\} \rightarrow 0$. 
\end{theorem}  

\vspace{.2cm}

\section{Facilitation of subsequent causal inference} \label{sec: use sas}

\def\Ms{{\mathcal {M}}}
\def\Ns{{\mathcal {N}}}
\def\comp{^{\mbox{\tiny{\sf C}}}}

We now explore how a consistent estimate of $\As\lo t$ can facilitate causal inference. In addition to estimating the average causal effect discussed below (\ref{eq: ate}) in \S 1, we focus on gaining insights into the dependence mechanism of $(Y(t),T,X)$ by studying their directed acyclic graph from $\As\lo t$. To link the two concepts rigorously, we adopt the Markov condition and the faithfulness assumption \citep[][\S2.1]{spirtes2010} throughout this section, which equate (\ref{as: vs ytx}) with the $d$-separation of $Y(t)$ and $T$ given $X\lo A$ in the directed acyclic graph. The faithfulness assumption also excludes the extreme case mentioned in \S 1, so that one of the active sets for $Y(t)|X$ and $T|X$, denoted by $A\lo {Y(t)}$ and $A\lo T$, respectively, must be a proper subset of $\{1,\ldots,p\}$. For ease of presentation, we assume $\As\lo t$ known throughout this section, and we leave the  review of the directed acyclic graph and the related concepts (e.g. path, fork, descendant, and $d$-separation), as well as the details about its examples used in this section, to the Appendix.    
%This determines how convincing our motivation is.

Recall from \S 1 that, when the average causal effect is of interest, the locally minimal sets are intuitively the optimal choices among all in $\As\lo t$ due to their locally minimal cardinalities. Because these sets may have different cardinalities, e.g. if we replace $X\lo 1$ and $X\lo 2$ in Figure \ref{fig: intro} with some larger $X\lo A$ and $X\lo B$, it is necessary to know all the locally minimal sets in $\As\lo t$ to find the optimal choice for the subsequent nonparametric estimation of the average causal effect. This illustrates the gain of knowing $\As\lo t$ in addressing the dimensionality issue for the causal effect estimation without parametric modeling.  

In regard of understanding the dependence mechanism between $Y(t)$ and $T$, $\As\lo t$ also provides unique information beyond the existing literature reviewed in \S 1. First, as illustrated in Figure \ref{fig: intro}, some sets in $\As\lo t$ (e.g. $\{3\}$) can index variables of $X$ that only have indirect effects on $Y(t)$ and $T$ through the rest of $X$. These variables cannot be detected by the existing methods that model $Y(t)|X$ and $T|X$, but they are potentially the fundamental cause that connects $Y(t)$ and $T$ and thus may attract the researchers' interest. To explore more from $\As\lo t$, we next study the bond between the locally minimal sets in $\As\lo t$ and $A\lo {Y(t)} \cap A\lo T$, i.e. that formed by the variables of $X$ who directly affect both $Y(t)$ and $T$. The key is the definition of $d$-separation and that, under the Markov condition and the faithfulness assumption, $A\lo {Y(t)} \cap A\lo T$ exactly indexes the forks between $Y(t)$ and $T$ in the directed acyclic graph.

\vspace{.2cm}

\begin{proposition} \label{prop: global min set}
Under the Markov condition and the faithfulness assumption, we have: 
($i$) $A\lo {Y(t)} \cap A\lo T$ is the intersection of all the locally minimal sets in $\As\lo t$; 
($i\!i$) $A\lo {Y(t)} \cap A\lo T \in \As\lo t$ if and only if $\As\lo t$ has the unique minimal set, in which case $A\lo {Y(t)} \cap A\lo T$ itself is the minimal set in $\As\lo t$.
\end{proposition}
%each $A \in \As\lo t$ must be a superset of $A\lo {Y(t)} \cap A\lo T$; conversely, let $A\lo {0, t}$ be the intersection of $A \in \As\lo t$, then $A\lo 0$ must 
% the nesting property holds for $\As$, that is, if $A \in \As\lo t$, then $B \in \As\lo t$ for any $B \supset A$.

\vspace{.2cm}

\begin{figure}[htbp]
	\centering
	\begin{subfigure}{0.23\linewidth}
		\centerline{
			\begin{tikzpicture}
				\node (1) at (0,1) {$X\lo 1$};
				\node (2) at (1.4,1) {$T$};
				\node (3) at (-1.4,0) {$X\lo 2$};
				\node (4) at (-1.4,1) {$Y(t)$};
				\node (5) at (0,0) {$X\lo 3$};
				\node (6) at (1.4,0) {$X\lo 4$};
    
				\path (1) edge  (2);
				\path (1) edge (4);
				\path (3) edge (4);
				\path (6) edge (2);
				\path (3) edge (5);
				\path (6) edge (5);
			\end{tikzpicture}
		}
		\caption{}
		\label{fig: unique min} 
	\end{subfigure}
	\hfill
	\begin{subfigure}{0.23\linewidth}
		\centerline{
			\begin{tikzpicture}
				\node (1) at (0,1) {$X\lo 1$};
				\node (2) at (-1.4,0) {$X\lo 2$};
				\node (3) at (0,0) {$X\lo 3$};
				\node (4) at (1.4,0) {$X\lo 4$};
				\node (5) at (-1.4,1) {$Y(t)$};
				\node (6) at (1.4,1) {$T$};
    
				\path (1) edge (5);
				\path (2) edge (5);
				\path (2) edge (3);
				\path (4) edge (3);
				\path (3) edge (1);
				\path (4) edge (6);
			\end{tikzpicture}
		}
		\caption{}
		\label{fig: ct ok cts ok}
	\end{subfigure}
 \hfill
	\begin{subfigure}{0.24\linewidth}
		\centerline{
			\begin{tikzpicture}
				\node (1) at (0,1) {$X\lo 1$};
				\node (2) at (-1.4,0) {$X\lo 2$};
				\node (3) at (0,0) {$X\lo 3$};
				\node (4) at (1.4,0) {$X\lo 4$};
				\node (5) at (-1.4,1) {$Y(t)$};
				\node (6) at (1.4,1) {$T$};
    
				\path (1) edge (5);
				\path (2) edge (5);
				\path (2) edge (3);
				\path (4) edge (3);
				\path (3) edge (5);
				\path (4) edge (6);
				\path (1) edge (6);
			\end{tikzpicture}
		}
		\caption{}
		\label{fig: ct cts both empty}
	\end{subfigure}
     \hfill
		\begin{subfigure}{0.24\linewidth}
		\centerline{
			\begin{tikzpicture}
				\node (1) at (0,1) {$X\lo 1$};
				\node (2) at (-1.4,0) {$X\lo 2$};
				\node (3) at (0,0) {$X\lo 3$};
				\node (4) at (1.4,0) {$X\lo 4$};
				\node (5) at (-1.4,1) {$Y(t)$};
				\node (6) at (1.4,1) {$T$};
    
				\path (1) edge (5);
				\path (2) edge (5);
				\path (2) edge (3);
				\path (4) edge (3);
				\path (3) edge (1);
				\path (4) edge (6);
				\path (1) edge (6);
			\end{tikzpicture}
		}
		\caption{}
		\label{fig: ct ok cts not}
	\end{subfigure}	
 \\[\baselineskip]
	\begin{subfigure}{0.35\linewidth}
		\centerline{
			\begin{tikzpicture}
				\node (1) at (-1.25,1) {$X\lo 1$};
				\node (2) at (0,1) {$X\lo 2$};
				\node (3) at (1.25,1) {$X\lo 3$};
				\node (4) at (-1.25,0) {$X\lo 4$};
				\node (5) at (0,0) {$X\lo 5$};
				\node (6) at (1.25,0) {$X\lo 6$};
				\node (7) at (-2.5,0.5) {$Y(t)$};
				\node (8) at (2.25,0.5) {$T$};
    
				\path (1) edge (7);
				\path (1) edge (2);
				\path (2) edge (5);
				\path (2) edge (4);
				\path (3) edge (2);
				\path (3) edge (8);
				\path (4) edge (7);
				\path (6) edge (5);
				\path (6) edge (8);
			\end{tikzpicture}
		}
		\caption{}
		\label{fig: complex}
	\end{subfigure}
    \hfill
	\begin{subfigure}{0.35\linewidth}
		\centerline{
			\begin{tikzpicture}
				\node (1) at (-0.5,1) {$X\lo 1$};
				\node (2) at (0.8,1) {$X\lo 2$};
				\node (3) at (-0.2,0) {$X\lo 3$};
				\node (4) at (-1.5,0) {$X\lo 4$};
				\node (5) at (1.1,0) {$X\lo 5$};
				\node (6) at (-2.25,1) {$Y(t)$};
				\node (7) at (2,1) {$T$};
    
				\path (1) edge (6);
				\path (1) edge (2);
				\path (2) edge (7);
				\path (3) edge (1);
				\path (4) edge (3);
				\path (5) edge (3);
				\path (4) edge (6);
				\path (5) edge (7);
			\end{tikzpicture}
		}
		\caption{}
		\label{fig: complex two}
	\end{subfigure}
    \hfill
		\begin{subfigure}{0.27\linewidth}
		\centerline{
			\begin{tikzpicture}
				\node (1) at (0,1) {$X\lo 1$};
				\node (2) at (-1.5,0) {$X\lo 2$};
				\node (3) at (0,0) {$X\lo 3$};
				\node (4) at (1.5,0) {$X\lo 4$};
				\node (5) at (-1.5,1) {$Y(t)$};
				\node (6) at (1.5,1) {$T$};
    
				\path (1) edge (5);
				\path (2) edge (5);
				\path (2) edge (3);
				\path (4) edge (3);
				\path (1) edge (3);
				\path (4) edge (6);
				\path (1) edge (6);
			\end{tikzpicture}
		}
		\caption{}
		\label{fig: ct ok cts not sub}
	\end{subfigure}	
	\centering
	\caption{Examples of directed acyclic graph of $(Y(t),T,X)$ with different $\As\lo t$'s.}
\end{figure}
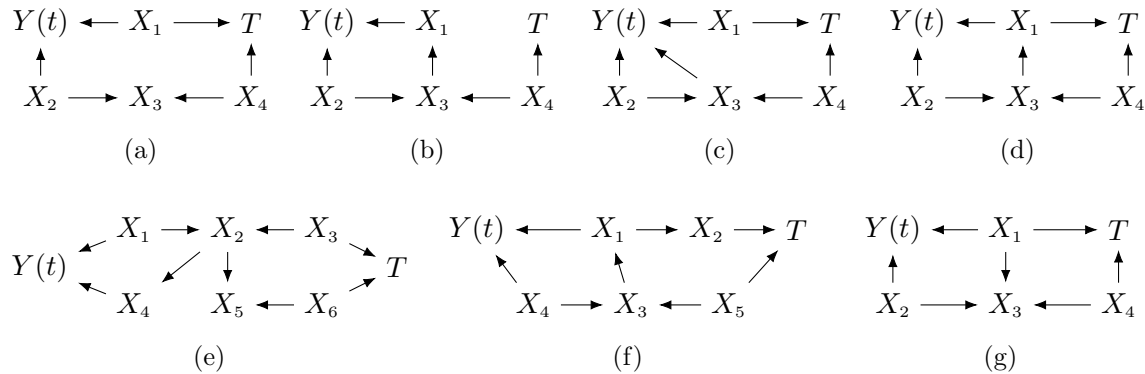

By Proposition \ref{prop: global min set}($i$), knowing $\As\lo t$ means knowing all the forks between $Y(t)$ and $T$. See Figure \ref{fig: unique min}, \ref{fig: ct cts both empty}, and \ref{fig: ct ok cts not} where the intersection of all the locally minimal sets in $\As\lo t$ is $\{1\}$ and $X\lo 1$ forms the only fork with $Y(t)$ and $T$, also Figure \ref{fig: ct ok cts ok}, \ref{fig: complex}, and \ref{fig: complex two} where this intersection is the empty set and no forks exist in the graph.
Due to the dependence between the variables of $X$, which may generate other paths between $Y(t)$ and $T$, $A\lo {Y(t)} \cap A\lo T$ alone is generally insufficient for the $d$-separation of $Y(t)$ and $T$, that is, $A\lo {Y(t)} \cap A\lo T \not\in \As\lo t$; see Figure \ref{fig: ct ok cts ok} where $A\lo {Y(t)} \cap A\lo T$ is the empty set but $Y(t)$ and $T$ are marginally $d$-connected. Proposition \ref{prop: global min set}($i\!i$) clarifies that $A\lo {Y(t)} \cap A\lo T \in \As\lo t$ if and only if $\As\lo t$ has the unique minimal set. In the language of $d$-separation, this holds if and only if every path between $Y(t)$ and $T$ that has no colliders must include some node that forms a fork between $Y(t)$ and $T$. The definition of a collider can be found in the next paragraph. In practice, this special case may often occur as it allows otherwise complexity of $\As\lo t$ resulted from the freedom of the paths with colliders. In particular, it is more general than the nesting property for $\As\lo t$ mentioned in \S 1; see Figure~\ref{fig: unique min} where $\As\lo t$ has the unique minimal set $\{1\}$ and the nesting property fails. 
%as $\As\lo t$ includes both $\{1,2,3\}$ and $\{1,3,4\}$ but not their intersection $\{1,3\}$.
%This can be rigorized using the results below, and will be revisited at the end of this section.
%From the discussions above, $\As\lo t$ can be complicated by the dependence within $X$ that induces indirect effects on $Y(t)$ and $T$. 
%requires that any $A, B \in \As\lo t$ implies $A\cap B \in \As\lo t$, which, by simple algebra

%\vspace{-0.25cm}

From the nesting of the sets in $\As\lo t$, we can also recover some important structures of the directed acyclic graph of $(Y(t), T,X)$. Namely, we can identify the types of nodes, particularly the aforementioned collider, that some variables of $X$ play in the graph. A collider means some $X\lo i$ who blocks a path between $Y(t)$ and $T$ via some $X\lo j \rightarrow X\lo i \leftarrow X\lo k$, which is unrelated to both $Y(t)$ and $T$ if only considering that path. Because a collider can meanwhile be a non-collider in some other path between $Y(t)$ and $T$ (see details below) and thus still related to $Y(t)$ and $T$, it ``carries with its unique considerations and challenges" to causal inference \citep[][\S2.3]{pearl2016}. To ease the presentation, we next abbreviate a path between $Y(t)$ and $T$ to a path, and abbreviate a set of a collider and its descendants to a collider, if no ambiguity is caused. For any $A \in \As\lo t$ and $i\in A$, we use $A \backslash \{i\}$ to denote the rest of $A$ if $i$ is removed. Let $\Ns\lo t \subset \As\lo t$ be the collection of $A \in \As\lo t$ such that all the supersets of $A$ are also included in $\As\lo t$. 
%For any $A\in \As\lo t$, we call $A$ a locally nested set if $\As\lo t$ also includes all the supersets of $A$

\vspace{.2cm}

\begin{proposition} \label{prop: what At tells}
Under the Markov condition and the faithfulness assumption, we have:

\noindent
($i$) for $i=1,\ldots, p$, if there exists $A \in \As\lo t$ such that $A \backslash \{i\} \not\in \As\lo t$, then $X\lo i$ is a non-collider in some path; if there exists $A \in \Ns\lo t$ such that $A \backslash \{i\} \in \As\lo t$ but $A \backslash \{i\} \not\in \Ns\lo t$, then $X\lo i$ is a non-collider in some path that has at least one collider;
%neither a collider nor a descendant of a collider

\noindent
($i\!i$) for any $B \in \Fs$, if there exists $A \in \As\lo t$ such that $A \cup B \not\in \As\lo t$ and $A \cup C \in \As\lo t$ for all $C \subsetneq B$, then $X\lo B$ consists of colliders in the same path. 
\end{proposition}

\vspace{.2cm}

An immediate corollary of Proposition \ref{prop: what At tells}($i$) is that any locally minimal set in $\As\lo t$ can only index the variables of $X$ that are non-colliders in some paths, which complies with both the interpretation of colliders and the intuition behind the local minimality. This also complements Proposition \ref{prop: global min set}. Let $C\lo t$ be the union of all $B \in \Fs$ in Proposition \ref{prop: what At tells}($i\!i$). Then $C\lo t$ must index colliders; see Figures \ref{fig: unique min}, \ref{fig: ct ok cts ok}, \ref{fig: ct ok cts not}, and \ref{fig: complex two} where it detects $X\lo 3$, and Figure \ref{fig: complex} where it detects $X\lo 2$ and $X\lo 5$. This detection however can be non-exhaustive; see Figure~\ref{fig: ct cts both empty} where $X\lo 3$ is a collider but $C\lo t$ is the empty set. Because the colliders that $C\lo t$ detects may as well be non-colliders in the other paths, e.g. $X\lo 3$ in Figures \ref{fig: ct ok cts ok}, \ref{fig: ct ok cts not}, and \ref{fig: complex two}, and $X\lo 2$ in Figure \ref{fig: complex}, we can refine $C\lo t$ by excluding its elements that satisfy Proposition \ref{prop: what At tells}($i$), and consider the rest as the candidate variables that only serve as colliders and thus redundant for any subsequent causal inference. This excludes $X\lo 3$ in Figures \ref{fig: ct ok cts ok} and \ref{fig: complex two}, and excludes $X\lo 2$ in Figure \ref{fig: complex}, but it fails to exclude $X\lo 3$ in Figure~\ref{fig: ct ok cts not}. 
%where it cannot  that is both a collider and a non-collider. 
%To refine the classification of nodes in $C\lo t$, 
%A precise interpretation of $C\lo t$ and $C\lo t\udex *$ requires strengthening the converse statements in Proposition \ref{prop: what At tells}, which we defer to future.
%More details about the examples in these two paragraphs that support the discussion above are deferred to the Appendix.
%, where we also provide more examples about $C\lo t$ and $C\lo t\udex *$ for different graphs. 

Due to the intrinsic limitation of $d$-separation \citep{spirtes2010}, $\As\lo t$ cannot completely determine the paths between $Y(t)$ and $T$ in general. For example, $\As\lo t$ is identical in Figures \ref{fig: ct ok cts not} and \ref{fig: ct ok cts not sub}. Nonetheless, it is still worth to explore more results towards this end besides those proposed above, as they help the researchers understand the fundamental connection between $Y(t)$ and $T$ via simple and semiparametric estimators of $\As\lo t$ rather than subtle parametric modeling \citep{spirtes2010}. This requires tremendous work and is deferred to future.   
%possible information that $\As\lo t$ can provide to the directed acyclic graph of $(Y(t),T,X)$

Finally, if the Markov condition and the faithfulness assumption hold and we partially know the directed acyclic graph of $(Y(t),T,X)$, then the results above can be used conversely to simplify the exhaustive searching process proposed in \S\ref{sec: method} and lighten the computational burden. These include: 
(a) by Proposition \ref{prop: global min set}($i\!i$), we only need to look into all the supersets of $A\lo {Y(t)} \cap A\lo T$;
(b) if some $X\lo i$ is known to be a collider in every path where it is present, that is, it is a common effect of other variables of $X$ and is otherwise unrelated to $Y(t)$ and $T$, then Proposition \ref{prop: what At tells}($i$) justifies that any $A \in \As\lo t$ implies $A \backslash \{i\} \in \As\lo t$; 
(c) if some $X\lo i$ is known to be a non-collider in every path where it is present, which occurs if it is not a common effect of any pairs of other variables of $X$, then Proposition \ref{prop: what At tells}($i\!i$) justifies that any $A\in \As\lo t$ implies $A \cup \{i\}\in \As\lo t$. 

\section{Simulation studies} \label{sec: sim}

We use the following five simulation examples to evaluate the proposed methods in recovering both $\As\lo t$ and its sub-collection of locally minimal sets, and in detecting colliders in $X$. The \texttt{R} code can be found in https://github.com/jun102653/confounder-selection. To examine the robustness of the proposed methods to Assumptions \ref{as: normal x|t} and \ref{as: gcopula x|t}, respectively, we generate $(X,T)$ under various distributions including the marginal normality of $X$ and a mixture of continuous and discrete components in $X$. For space limit, we set $p=10$ and defer $p=20$ to the Appendix. In the following, $B(\pi)$ denotes the Bernoulli distribution with mean $\exp(\pi)/\{1+\exp(\pi)\}$, $e\lo {1,2}$ denotes $(1,1,0,\ldots, 0)\trans \in \real\udex p$, and each $\zeta\lo i(\cdot)$ is fixed as $\Phi\inv (F\lo {2,3}(\cdot))$, $F\lo {2,3}(\cdot)$ being the cumulative distribution function of the F distribution with the degrees of freedom $(2,3)$. The random errors $\epsilon\lo t$ for $t=0,1$ and $\varepsilon\lo i$ for $i=1,\ldots, 5$ are generated from $N(0, .2)$. 
%except in Models 1-2 where $\epsilon\lo 0$ and $\epsilon\lo 1$ are generated from $N(0, 1)$.   
%For reference, \cite{de2011}'s approach is also included in the comparison, which again only detects two locally minimal sets in $\As\lo t$.

\begin{enumerate}
\item $Y(t)=(4+t)(X\lo 2+X\lo 3) + 2.2\epsilon\lo t$, $T \sim B(0)$, $X\lo {-4}|T \sim N(\mu\lo T, \Sigma\lo T)$, \\
$\mu\lo 1= .6 e\lo {1,2}$, $\mu\lo 0 = 0$, $\Sigma\lo 0=\Sigma\lo 1=.8 I\lo p$, $X\lo 4=1.5X\lo 3+X\lo 1+\varepsilon\lo 4$.

\vspace{.2cm}

\item $Y(t)=(9+t)\sin(X\lo 2)+ 9 (1-t) X\lo 3\udex 3 + 10 t \sin(X\lo 3) + 2.2\epsilon\lo t$, $T \sim B(0)$, \\
$X\lo {-4}|T \sim N(\mu\lo T, \Sigma\lo T)$,
$\mu\lo 1=.5 e\lo {1,2}$, $\mu\lo 0 = 0$, $\Sigma\lo 0=\Sigma\lo 1= .6 I\lo p, X\lo 4 = 2 X\lo 3 + 2 X\lo 1 + \varepsilon\lo 4$.

\vspace{.2cm}

\item $Y(t)= 2X\lo 6+ .4(1+t) X\lo p\udex 3 + 7X\lo 3/(.5+(X\lo 1+2)\udex 3)+ \epsilon\lo t$, 
$T \sim B(X\lo 2+X\lo 5)$, \\
$X\lo {4,8,9,10} \sim N(0, .6 I\lo 4)$, 
$X\lo {1,2,3,5} = X\lo 4 (2,2,2,2)\trans + \varepsilon\lo {1,2,3,5}$, $X\lo 6 \sim B(0)$, $X\lo 7 \sim B(0)$. 

\vspace{.2cm}

\item $Y(t)= X\lo 1 + (1.5-.5t) X\lo 2 + \sin(X\lo 3) + \epsilon\lo t$,
$T \sim B(0)$, \\
$\zeta(X)|T \sim N(0, \Sigma\lo T)$, $\Sigma\lo 0 =I\lo p$, $\Sigma\lo 1 = I\lo p$ except for $\Sigma\lo 1\udex {(1,2)} = \Sigma\lo 1\udex {(2,1)}=.5$.

\vspace{.2cm}

\item Same as Model $4$ but with $\epsilon\lo t$ in generating $Y(t)$ replaced by $10 \epsilon\lo t$.
\end{enumerate}

These models represent a variety of cases from simple to complex patterns and from weak to strong signals. The effect of $X$ on $Y(t)$ is linear in Model 1 and is complex otherwise, and is much weaker in Model 5 than in Model 4. The effect of $X$ on $T$ conveys a logistic regression in Models 1-3 and is more subtle in Models 4-5. In Models 1-2, $X\lo 4$ is a collider and detectable by $C\lo t$. In Model 3, $X\lo 4$ affects $Y(t)$ and $T$ indirectly and induces one of the three locally minimal sets in $\As\lo t$. Because $\As\lo t$ has a large cardinality in all the models, i.e. $448$, $448$, $736$, $256$, and $256$, respectively, its recovery is nontrivial especially in Model 5 with weak signals.

\def\Cs{\mathcal {C}}

Considering the symmetry of $X|T$ in Models 4-5, we use the sliced average variance estimator to construct $\M\lo T$ and $\M\lo {T|\zeta(X)}$ for these models, and we use the sliced inverse regression uniformly for the other candidate matrices. To measure the similarity between $\As\lo t$ and some $\widehat \As\lo t$, we use
\begin{align}\label{eq: rho and omega for At}
\rho\lo t \equiv \Cs(\widehat \As\lo t \cap \As\lo t)\, / \, \Cs(\As\lo t) \quad \mbox{and} \quad
\omega\lo t \equiv \Cs(\widehat \As\lo t \cap \As\lo t) \, / \, \Cs(\widehat \As\lo t),
\end{align}
where $\Cs(\cdot)$ denotes the cardinality of a collection. Both measures have range $(0, 1)$ with larger values indicating better similarity, and they complement each other as $\rho\lo t$ is sensitive to the number of falsely unselected sets in $\As\lo t$ and $\omega\lo t$ is sensitive to the number of falsely selected sets not in $\As\lo t$. We also report the probability that $\widehat \As\lo t$ includes all the locally minimal sets. To see whether $\widehat \As\lo t$ truly collects the collider set $C\lo t$, we use the rule introduced below Proposition \ref{prop: what At tells} to derive $\widehat C\lo t$ based on $\widehat \As\lo t$, and record $\Cs(\widehat C\lo t \cap C\lo t)$ and $\Cs(\widehat C\lo t \backslash C\lo t)$, the numbers of truly and falsely selected colliders. The average of these measures among $2000$ independent runs are recorded in Table \ref{table: sim}, where the sample size $n$ is set at $400$ and $800$ sequentially. For convenience, we call the proposed method that assume normal $X|T$ the normality estimator, and call that assume a Gaussian copula on $X|T$ the Gaussian copula estimator, and denote them by ``MN" and ``GC" in Table \ref{table: sim}.
%For convenience, we call the proposed method that assume the normality and a Gaussian copula on $X|T$ the normality estimator and the Gaussian copula estimator, respectively, and denote them by ``MN" and ``GC" in Table \ref{table: sim}.

\def\Ts{{\mathcal T}}
\def\Fs{{\mathcal F}}

\begin{table}[b]
\vspace*{-6pt}
\centering
\def\~{\hphantom{0}}
\begin{tabular*}{\columnwidth} {@{}l@{\extracolsep{\fill}}r@{\extracolsep{\fill}}r@{\extracolsep{\fill}}r@{\extracolsep{\fill}}r@{\extracolsep{\fill}}r@{\extracolsep{\fill}}r@{\extracolsep{\fill}}r@{\extracolsep{\fill}}r@{\extracolsep{\fill}}r@{\extracolsep{\fill}}r@{\extracolsep{\fill}}r@{}} \\
	\hline
\empty & \multicolumn{2}{c}{{Model 1}} & \multicolumn{2}{c}{{Model 2}} & \multicolumn{2}{c}{{Model 3}} & \multicolumn{2}{c}{{Model 4}} & \multicolumn{2}{c}{{Model 5}}  \\ \hline 
$n=400$ &\multicolumn{1}{c}{MN}  & \multicolumn{1}{c}{GC} & \multicolumn{1}{c}{MN}  & \multicolumn{1}{c}{GC}  & \multicolumn{1}{c}{MN} & \multicolumn{1}{c}{GC} & \multicolumn{1}{c}{MN}  & \multicolumn{1}{c}{GC} & \multicolumn{1}{c}{MN} & \multicolumn{1}{c}{GC}   \\ \hline
 %\hline \\ [-6pt]
	%\hline 
	$\rho\lo 0/\omega\lo 0$ & 92/94&	92/94&	91/96&	91/96&	100/74	&100/74&	23/78&	100/93	&23/77	&96/76 \\
	$\rho\lo 1/\omega\lo 1$ &  92/95&	92/94&	90/95&	89/95&	94/76&	96/75&	24/76&	100/89&	23/76&	92/73 \\		
	$\pi\lo 0 / \pi\lo 1$ & 80/79&	81/80&	75/74&	77/75&	100/93	&100/95&	10/11&	100/100&	10/10&	96/92
\\
	$\Ts\lo 0 / \Fs\lo 0$ & 45/31&	46/31&	53/19&	53/18&	$-/0$&	$-/0$&	$-/0$&	$-/2$&	$-/0$&	$-/6$\\
	$\Ts\lo 1 / \Fs\lo 1$ & 47/28&	46/32&	54/49&	55/47&	$-/4$&	$-/3$&	$-/0$&	$-/5$&	$-/0$&	$-/5$\\ \hline
$n=800$ &\multicolumn{1}{c}{MN}  & \multicolumn{1}{c}{GC} & \multicolumn{1}{c}{MN}  & \multicolumn{1}{c}{GC}  & \multicolumn{1}{c}{MN} & \multicolumn{1}{c}{GC} & \multicolumn{1}{c}{MN}  & \multicolumn{1}{c}{GC} & \multicolumn{1}{c}{MN} & \multicolumn{1}{c}{GC}   \\ \hline
$\rho\lo 0/\omega\lo 0$ & 96/98&	96/98&	95/99&	95/99&	100/74&	100/74&	26/76&	100/100& 26/76&	100/98 \\
	$\rho\lo 1/\omega\lo 1$ &  96/98&	96/98&	93/98&	94/98&	99/74	&100/74&	28/75&	100/100&	26/74&	100/96\\		
	$\pi\lo 0 /\pi\lo 1$ & 88/88&	89/88&	84/81&	87/82&	100/99&	100/100&	13/14&	100/100&	13/12&	100/100\\
	$\Ts\lo 0 / \Fs\lo 0$ & 76/4&	75/5&	77/3&	79/2&	$-/0$&	$-/0$&	$-/0$&	$-/0$&	$-/0$&	$-/0$
 \\
$\Ts\lo 1 / \Fs\lo 1$ &  75/6&	75/6&	75/21&	76/22&	$-/1$&	$-/0$&	$-/0$&	$-/0$&	$-/0$&	$-/2$ \\ \hline
 %\hline
\end{tabular*}%\vskip18pt
\caption{The performance of the proposed methods in the simulation models: ``MN" stands for the normality estimator, and ``GC" for the Gaussian copula estimator; for $t=0, 1$, $\rho\lo t$ and $\omega\lo t$ measure the accuracy of recovering $\As\lo t$ and are defined in (\ref{eq: rho and omega for At}), $\pi\lo t$ indicates whether all the locally minimal sets in $\As\lo t$ are detected, and $\Ts\lo t$ and $\Fs\lo t$ denote the numbers of truly and falsely selected colliders, $\Ts\lo t$ unavailable for Models 3-5 where no colliders exist; these measures are averaged based on $2000$ independent runs and then multiplied by $100$.}
\label{table: sim}
\end{table}

From Table \ref{table: sim}, the two methods consistently recover both $\As\lo 0$ and $\As\lo 1$ in Models 1-3, where their assumptions on $(X,T)$ are violated in Model 3. The Gaussian copula estimator also consistently recovers $\As\lo 0$ and $\As\lo t$ in Model 4, and its performance is only slightly compromised in Model 5 where the signal is weaker. The inconsistency of the normality estimator in Models 4-5 meets the theoretical anticipation, although the large $\omega\lo 0$ and $\omega\lo 1$ indicate that it only under-selects $\As\lo 0$ and $\As\lo t$, that is, most elements in the corresponding $\widehat \As\lo 0$ and $\widehat \As\lo 1$ are still sufficient adjustment sets. 
The consistency of the proposed methods in recovering the minimal sets follows that in recovering $\As\lo 0$ and $\As\lo 1$, which particularly suggests their capability in capturing the indirect effects of $X$ in Model 3. The methods are less accurate in detecting the colliders, but their improved performance when $n$ increases to $800$ still indicates their asymptotic consistency in this respect. The otherwise performances of the proposed methods are also enhanced when $n$ grows from $400$ to $800$, except for the normality estimator in Models 4-5.

%\newpage
\section{Application}
\label{s:application}

We apply the proposed work to a study of whether the mother smoking during pregnancy causes the infant's low birth weight \citep{kramer1987}. The dataset available at http://www.stata-press.com/data/r13/cattaneo2.dta. was analyzed in \cite{almond2005}, \cite{da2008}, and \cite{lee2017} using a program evaluation approach and a propensity score matching method, etc..   Following \cite{lee2017}, we restrict the sample to the white and non-Hispanic mothers, which include $3754$ subjects. The outcome variable is the infant's birth weight measured in grams, and the predictor includes the indicator of alcohol consumption during pregnancy ($X\lo 1$), the indicator of a previous dead newborn ($X\lo 2$), the mother’s age ($X\lo 3$), the mother’s educational attainment ($X\lo 4$), the number of prenatal care visits ($X\lo 5$), the indicator of the newborn being the first baby ($X\lo 6$), and the indicator for the first prenatal visit in the first trimester ($X\lo 7$). The ignorability assumption (\ref{as: ignorability}) was justified in \cite{lee2017}, and the common support assumption (\ref{as: common supp}) is supported by an omitted exploratory data analysis. 

\begin{figure}[htbp]
\centerline{\includegraphics[width=300pt,height=140pt]{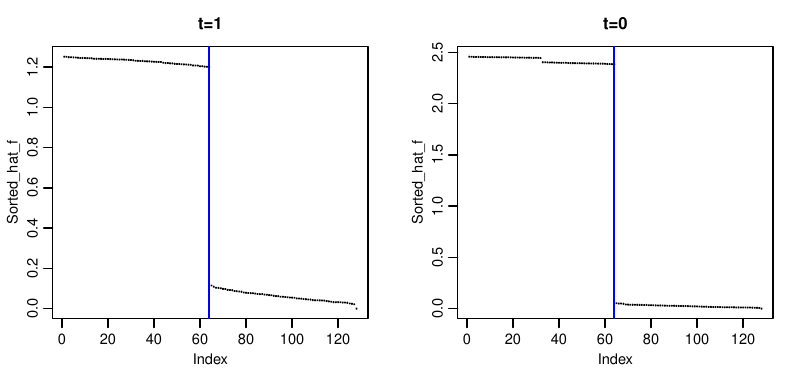}}
\caption{The scree plots of $\widehat f\lo t(A\lo {k}\udex {(t)})$'s for the birth weight dataset: the left panel is for $t=1$ and the right panel is for $t=0$; the vertical line tells the cutoff of the tail for each plot determined by the ridge ratio method.}
\label{fig: data}
\end{figure}

Under the concern of complex $X|T$, we apply the proposed Gaussian copula estimator with both $\M\lo {T|\zeta(X)}$  and $\M\lo {Y(t)|\zeta(X)}$ from directional regression. As seen in the scree plots of $\widehat f\lo 0 (A\lo k\udex {(0)})$'s and $\widehat f\lo 1 (A\lo k\udex {(1)})$'s in Figure \ref{fig: data}, both $\widehat \As\lo 0$ and $\widehat \As\lo 1$ include the unique minimal set $\{1\}$ and all of its supersets. This means that the mother's alcohol consumption during pregnancy satisfies (\ref{as: vs ytx}) alone for both $t=0, 1$, and, by Proposition \ref{prop: global min set}, it is the only factor associated with both the mother's smoking and the infant's potential birth weights had the mother smoked and had her not smoked. 

Using the mother's alcohol consumption as the univariate predictor, we apply the one-to-one matching method in the \texttt{R} package \texttt{Matching} to estimate the average causal effect. The estimate is $-281$ grams, with the bootstrap standard deviation equal to $58.2$ grams. This suggests a significant causal effect of the mother's smoking on the infant's low birth weight, which is consistent with the findings in the previous studies \citep{kramer1987,almond2005,lee2017}. Applying the same method to the original data with the ambient predictor, the estimate of the average causal effect is similarly $-263$ grams, but its bootstrap standard deviation is raised to $103$ grams, resulting in a less significant conclusion. This illustrates the gain of using small sufficient adjustment sets in sharpening the causal effect estimation. 

% Acknowledgements and Disclosure of Funding should go at the end, before appendices and references

\acks{Wei Luo's research was supported in part by the National Science Foundation
of China (12222117, 12131006). Lixing Zhu's research was supported in part by the National Science Foundation of China (12131006).}


\newpage

\appendix

\section{Consistency of $\widehat \As\lo t$ under the high-dimensional settings} 
\label{sec: hd}

% Note: in this sample, the section number is hard-coded in. Following
% proper LaTeX conventions, it should properly be coded as a reference:

%In this appendix we prove the following theorem from
%Section~\ref{sec:textree-generalization}:

When $p$ diverges with $n$, $\Sigma$, $\M\lo {Y(t)}$, and $\M\lo {T}$ are sequences of matrices with growing dimensions and dynamic entries. Thus, we must regulate the signal strength of the data through these matrices, to preclude weak effects that are infeasible to detect in the sample level. Let $\lambda\lo \min (\cdot)$ denote the minimal nonzero singular value of a matrix, and let ${\mathcal M}$ be the collection of matrix sequences $(\M\lo p)$ with $\|\M\lo p\| = O(1)$ and $\lambda\lo \min\inv (\M\lo p) = O(1)$. Without loss of generality, suppose $\beta\lo {Y(t)}$ and $\beta\lo T$ that span $\cs {Y(t)}$ and $\cs T$, respectively, are sequences of semi-orthogonal matrices.
%For the consistency of $\widehat \As\lo t$ w

%the tail of the scree plot formed by the rearranged $\widehat f\lo t(\cdot)$, and propose
%, with $p$ omitted from their subscripts for simplicity
% the possibility of equal values being ignored without loss of generality
%including the BIC-type methods \citep{zhu2006sliced}, the sequential testing procedure \citep{jolliffe2002,zhu2006}, and the thresholding double ridge ratio method \citep{zhu2020}, etc.  

\vspace{0.2cm}

\begin{assumption} \label{as: signal strength}
($i$) $\Sigma\udex {(0)}, \Sigma\udex {(1)}, \M\lo {Y(t)}, \M\lo T \in {\mathcal M}$; ($i\!i$) There exists $c>0$ such that, for $s=0, 1$ and all the large $p$, $\min\{ E[ \|\cov(\beta\lo {Y(t)}\trans X, \beta\lo T\trans X | X\lo A, T=s)\|]: A \not\in \As\lo t\} > c$.  
\end{assumption}

\vspace{0.2cm}

\noindent
Part ($i$) of this assumption requires that the components of $X$ are not ill-correlated, and that each direction in $\cs {Y(t)}$ (and likewise in $\cs T$) has uniformly non-vanishing effect on $Y(t)$, both of which are commonly adopted in the relative literature \citep{zhu2010cse,qian2019,tan2019}. Part ($i\!i$) of Assumption \ref{as: signal strength} regulates $\cs {Y(t)}$ and $\cs T$ to preclude any subtle ``local alternative hypothesis" whenever the statements in (14) fail. To incorporate the divergence of $p$, we write $c\lo n$ in (16) as $c\lo {n,p}$, which can vary with $p$. 
%These together imply a non-vanishing $f\lo t(A)$ whenever $A \not\in \As\lo t$.   

\vspace{0.2cm}

\begin{theorem} \label{thm: hd normal x|t}
Suppose $\| (\widehat \M\lo {Y(t)}, \widehat{\M}\lo T, \widehat \Sigma\udex {(0)}, \widehat \Sigma\udex {(1)} ) - (\M\lo {Y(t)}, \M\lo T, \Sigma\udex {(0)}, \Sigma\udex{(1)}) \| = \bop (r\lo {n,p})$ for some $r\lo {n,p} = o(1)$. Under the ignorability assumption (\ref{as: ignorability}), the common support condition (\ref{as: common supp}), Assumption \ref{as: normal x|t}, and Assumption \ref{as: signal strength}, we set $c\lo {n,p}$ in (\ref{eq: taut}) such that $\max\{r\lo {n,p} / c\lo {n,p}, c\lo {n,p}\}= o(1)$, then $\widehat \As\lo t$ coincides with $\As\lo t$ with probability converging to one. %with probability converging to one.
\end{theorem}

\vspace{0.2cm}

The lower bound $r\lo {n,p}$ for $c\lo {n,p}$ in Theorem \ref{thm: hd normal x|t} depends on the diverging rate of $p$ with respect to $n$ as well as the specific inverse regression methods that induce $\widehat \M\lo {Y(t)}$ and $\widehat \M\lo T$. For example, when $ n\inv p \log p \rightarrow 0$, \cite{wainwright2019} justified $\|\widehat \Sigma - \Sigma \| = \bop \{(p \log p)\udex {1/2} n\udex {-1/2}\}$, and \cite{qian2019} justified that both $\| \widehat \M\lo {Y(t)}-\M\lo {Y(t)} \|$ and $\| \widehat{\M}\lo T - \M\lo T \|$ are also $\bop \{(p \log p)\udex {1/2} \\ n\udex {-1/2}\}$ for the sliced inverse regression, assuming that both $\cs {Y(t)}$ and $\cs T$ are sparse and have fixed dimensions as $p$ diverges; these together imply $r\lo {n,p} = (p \log p)\udex {1/2} n\udex {-1/2}$, that is, \\ $c\lo {n,p} n\udex {1/2} /(p\log p)\udex {1/2} \rightarrow \infty$ for the consistency of $\widehat \As\lo t$. 

\section{Review of the directed acyclic graph}
\label{sec: dag}

\newcommand\cone[1]{{\mathcal {N}} \{#1\}}

We now briefly review the directed acyclic graph commonly used to depict the dependence mechanism of a set of random variables. Each random variable is represented by a node (or called a vertex) in the graph, and their overall dependence is represented by the set of directed edges between the nodes. For example, for any pair of random variables $(R,W)$, $R \rightarrow W$ means that $R$ is a cause of $W$ and is called an ascendant (or parent) of $W$, and $W$ is called a descendant (or child) of $R$. The meaningfulness of these edges requires the local Markov property that there exists a joint distribution of the nodes such that each node is conditionally independent of its non-descendants given its ascendants. Generally, a directed acyclic graph determines the corresponding probabilistic dependence between its nodes, but conversely the probabilistic dependence of a set of variables can only determine the directed acyclic graph up to its equivalence class \citep{pearl2016}.

A path in the directed acyclic graph is defined as a set of edges such that each node it involves is visited only once. Any path can be decomposed into a set of triplets $(R,W,Z)$. The case $R \rightarrow W \rightarrow Z$ is called a chain, and it occurs if $R$ is a direct cause of $W$ and $W$ is a direct cause of $Z$. The case $R\leftarrow W \rightarrow Z$ is called a fork, and it occurs if $W$ is a common cause of both $R$ and $Z$. Both of these mean $R\indep Z \mid W$ and thus are indistinguishable by the probablisitc dependence of $(R,W,Z)$, The case $R \rightarrow W \leftarrow Z$ is called a collider, and it occurs if $W$ is a common effect of both $R$ and $Z$. For ease of presentation, we also call $W$ a collider in this case. A path is called blocked by a set of nodes if this set of nodes include at least one of the non-colliders of the path, or if this set of nodes do not include at least one of the colliders (together with its descendants) of the path. A pair of nodes $R$ and $Z$ are called conditionally $d$-separated given a set of nodes $W$ if every path between $R$ and $Z$, i.e. with these two as the ending nodes, is blocked by $W$. Namely, this occurs if and only if both of the following two statements hold:

\vspace{0.2cm}

\noindent
{\it ($i$) for each path between $R$ and $Z$ that has no colliders, $W$ includes at least one of its nodes;}

\vspace{0.2cm}

\noindent
{\it ($i\!i$) for each path between $R$ and $Z$ that has an non-empty set of colliders $U$, where each element of $U$ also represents its deccendants for simplicity, $W$ is either not a superset of $U$ or includes at least one of the non-colliders in this path.}

\vspace{0.2cm}

\noindent
Under the Markov condition and the faithfulness assumption \citep[][\S 2.1]{spirtes2010}, the conditional $d$-separation between any $R$ and $Z$ given $W$ is equivalent to the probabilistic independence $R \indep Z | W$. Thus, the definition of $d$-separation above will be used to provide $\As\lo t$ for the examples of directed acyclic graph and to prove the relative propositions listed in the main text. The former is presented in the rest of this section, and the latter is deferred to Appendix \ref{sec: proof} later. For simplicity, we write any $\{i\lo 1,\ldots,i\lo k\} \in \Fs$ as $(i\lo 1 \ldots i\lo k)$ throughout this section, and use $\cone{i\lo 1 \ldots i\lo k}$ to denote the collection of all the supersets of $(i\lo 1 \ldots i\lo k)$. The former should not cause ambiguity with the labels of the equations in the main text, which only appear in the other sections of the appendix. 
%, both commonly adopted in the literature \citep{de2011,haggstrom2018},

%and indicates that all conditional (in-)dependence relations can be read-off from the directed acyclic graph using the d-separation rule (Glymour et al., 2016). Specifically, the conditional independence between variables $A$ and $B$ given variables $C$ can be read off from the graph by checking whether the nodes in $C$ block every path between nodes $A$ and $B$. A path between $A$ and $B$ is blocked by $C$ if (a) it contains a noncollider in $C$ or (b) it contains a collider such that neither the collider itself
%nor any descendant thereof are in $C$. The special case of marginal independence between variables $A$ and $B$ can be seen by checking whether the empty set blocks every path between $A$ and $B$. This is only possible if there is no path between $A$ and $B$ or every path contains a collider.

%\vspace{0.2cm}

%\noindent
The following causal graph is the example in Section \ref{sec: intro} of the main text.

\vspace{-0.35cm}

\begin{figure}[htbp]
	\centering
		\centerline{
			
			\begin{tikzpicture}
				\node (1) at (-1.5,1) {$X\lo 1$};
				\node (3) at (0,1) {$X\lo 3$};
				\node (2) at (1.5,1) {$X\lo 2$};
				\node (4) at (-1.5,0) {$X\lo 4$};
				\node (5) at (0,0) {$X\lo 5$};
				\node (6) at (1.5,0) {$X\lo 6$};
				\node (7) at (-3,1) {$Y(t)$};
				\node (8) at (3,1) {$T$};
    
				\path (1) edge (7);
				\path (4) edge (7);
				\path (3) edge (1);
				\path (3) edge (2);
				\path (4) edge (5);
				\path (6) edge (5);
				\path (2) edge (8);
				\path (6) edge (8);
			\end{tikzpicture}
		}
		%\caption{\label{fig3a}(a)}

	\centering
\end{figure}

\vspace{-0.30cm}

\noindent
It contains two paths between $Y(t)$ and $T$: $Y(t) \leftarrow X\lo 1 \leftarrow X\lo 3 \rightarrow X\lo 2 \rightarrow T$ and $Y(t) \leftarrow X\lo 4 \rightarrow X\lo 5 \leftarrow X\lo 6 \rightarrow T$, where the latter include a collider $X\lo 5$. Thus, the $d$-separation of $Y(t)$ and $T$ given $A \in \Fs$ is equivalent to that $A$ indexes at least one variable in the first path and meanwhile either indexes $X\lo 4$ or $X\lo 6$ or does not index $X\lo 5$ in the second path. This means 
\begin{align*}
\As\lo t =  \{(1), (2), (3), (12), (13), (23), (123)\} \cup \left[\cup\lo {i=1,2,3, j=4,6}\cone{ij}\right], 
\end{align*}
with the locally minimal sets being $(1)$, $(2)$, and $(3)$. Note that $(1)\cap (2) = \emptyset$ and there is no unique minimal set in $\As\lo t$ in this case, which comply with Proposition \ref{prop: global min set} of the main text.

\medskip

The following graph is the example in Figure \ref{fig: unique min} of the main text.

%\vspace{-0.05cm}

\begin{figure}[htbp]
	\centering
		\centerline{
			
			\begin{tikzpicture}
				\node (1) at (0,1) {$X_1$};
				\node (2) at (2,1) {$T$};
				\node (3) at (-2,0) {$X_2$};
				\node (4) at (-2,1) {$Y(t)$};
				\node (5) at (0,0) {$X_3$};
				\node (6) at (2,0) {$X_4$};
    
				\path (1) edge  (2);
				\path (1) edge (4);
				\path (3) edge (4);
				\path (6) edge (2);
				\path (3) edge (5);
				\path (6) edge (5);
			\end{tikzpicture}
}
		%\caption{\label{fig3a}(a)}

	\centering
	
\end{figure}

%\vspace{-0.30cm}

\noindent
It contains two paths between $Y(t)$ and $T$: $Y(t) \leftarrow X\lo 1 \rightarrow T$ and  $Y(t) \leftarrow X\lo 2 \rightarrow X\lo 3 \leftarrow X\lo 4 \rightarrow T$, the latter including a collider. Thus, the $d$-separation of $Y(t)$ and $T$ given $A \in \Fs$ is equivalent to that, first, $A$ includes $(1)$, second, $A$ either does not index $X\lo 3$ or indexes at least one of the other nodes in the second path. This means 
\begin{align*}
\As\lo t = \{(1)\} \cup \cone{12} \cup  \cone{14}.
\end{align*}
Hence, $\As\lo t$ has the unique minimal set $(1)$ but does not satisfy the nesting property. Since $(1) \in \As\lo t$ and $(13) \not\in \As\lo t$, $i=3$ satisfies Proposition \ref{prop: what At tells}($i\!i$), so we have $C\lo t = (3)$.

%\medskip

The following graph is the example in Figure \ref{fig: ct ok cts ok} of the main text.

\vspace{-0.30cm}

\begin{figure}[htbp]
	\centering
		\centerline{
			
			\begin{tikzpicture}
				\node (1) at (0,1) {$X_1$};
				\node (2) at (-2,0) {$X_2$};
				\node (3) at (0,0) {$X_3$};
				\node (4) at (2,0) {$X_4$};
				\node (5) at (-2,1) {$Y(t)$};
				\node (6) at (2,1) {$T$};
    
				\path (1) edge (5);
				\path (2) edge (5);
				\path (2) edge (3);
				\path (4) edge (3);
				\path (3) edge (1);
				\path (4) edge (6);
			\end{tikzpicture}
}
		%\caption{\label{fig3a}(a)}

	\centering
	
\end{figure}

\vspace{-0.30cm}

\noindent
It contains two paths between $Y(t)$ and $T$: $Y(t) \leftarrow X\lo 1 \leftarrow X\lo 3 \leftarrow X\lo 4 \rightarrow T$ and $Y(t) \leftarrow X\lo 2 \rightarrow X\lo 3 \leftarrow X\lo 4 \rightarrow T$, the latter including a collider. Thus, the $d$-separation of $Y(t)$ and $T$ given $A \in \Fs$ is equivalent to that, first, $A$ indexes at least one variable in the first path, second, $A$ either does not index $X\lo 3$ or indexes at least one of the other nodes in the second path. This means 
\begin{align*}
\As\lo t = \cone{4} \cup \{(1), (12), (23), (123)\}.
\end{align*}
Consequently, the intersection of all the local minimal sets in $\As\lo t$ is the empty set, and we have $C\lo t = (3)$ as only $i=3$ satisfies Proposition \ref{prop: what At tells}($i\!i$) with $(1) \in \As\lo t$ and $(13) \not\in \As\lo t$. Because $(23) \in \As\lo t$ and $(2) \not\in \As\lo t$, $i=3$ also satisfies Proposition \ref{prop: what At tells}($i$) and thus is removed from the refined $C\lo t$.

%%% This is (dddddddddddddddddddddddddddddddd)

%\vspace{0.2cm}

The following graph is the example in Figure \ref{fig: ct cts both empty}  of the main text.

\vspace{-0.30cm}

\begin{figure}[htbp]
	\centering
		\centerline{
			
			\begin{tikzpicture}
				\node (1) at (0,1) {$X_1$};
				\node (2) at (-2,0) {$X_2$};
				\node (3) at (0,0) {$X_3$};
				\node (4) at (2,0) {$X_4$};
				\node (5) at (-2,1) {$Y(t)$};
				\node (6) at (2,1) {$T$};
    
				\path (1) edge (5);
				\path (2) edge (5);
				\path (2) edge (3);
				\path (4) edge (3);
				\path (3) edge (5);
				\path (4) edge (6);
				\path (1) edge (6);
			\end{tikzpicture}
}
		%\caption{\label{fig3a}(a)}

	\centering
	
\end{figure}

\vspace{-0.30cm}

\noindent
It contains three paths between $Y(t)$ and $T$: $Y(t) \leftarrow X\lo 1 \rightarrow T$,  $Y(t) \leftarrow X\lo 3 \leftarrow X\lo 4 \rightarrow T$, and $Y(t) \leftarrow X\lo 2 \rightarrow X\lo 3 \leftarrow X\lo 4 \rightarrow T$, the latter including a collider. Thus, the $d$-separation of $Y(t)$ and $T$ given $A \in \Fs$ is equivalent to that, first, $A$ indexes at least one variable in each of the first two paths, second, $A$ either does not index $X\lo 3$ or indexes at least one of the other nodes in the last path. This means 
\begin{align*}
\As\lo t = \cone{14} \cup \cone {123}.
\end{align*}
Consequently, the intersection of all the locally minimal sets in $\As\lo t$ is $(1)$, and we have $C\lo t = \emptyset$ as no elements in $(1234)$, including $i=3$, satisfy Proposition \ref{prop: what At tells} ($i\!i$). 
%Since $i=3$ satisfies Proposition \ref{prop: what At tells}($i$) with $(123) \in \As\lo t$ and $(12) \not\in \As\lo t$, it is excluded from the refined $C\lo t$.

%\vspace{0.2cm}

The following graph is the example in Figure \ref{fig: ct ok cts not} of the main text.

%\vspace{-0.30cm}

\begin{figure}[htbp]
	\centering
		\centerline{
			
			\begin{tikzpicture}
				\node (1) at (0,1) {$X_1$};
				\node (2) at (-2,0) {$X_2$};
				\node (3) at (0,0) {$X_3$};
				\node (4) at (2,0) {$X_4$};
				\node (5) at (-2,1) {$Y(t)$};
				\node (6) at (2,1) {$T$};
    
				\path (1) edge (5);
				\path (2) edge (5);
				\path (2) edge (3);
				\path (4) edge (3);
				\path (3) edge (1);
				\path (4) edge (6);
				\path (1) edge (6);
			\end{tikzpicture}
}
		%\caption{\label{fig3a}(a)}

	\centering
	
\end{figure}

%\vspace{-0.30cm}

\noindent
It contains four paths between $Y(t)$ and $T$: $Y(t) \leftarrow X\lo 1 \rightarrow T$,  $Y(t) \leftarrow X\lo 1 \leftarrow X\lo 3 \leftarrow X\lo 4 \rightarrow T$, $Y(t) \leftarrow X\lo 2 \rightarrow X\lo 3 \rightarrow X\lo 1 \rightarrow T$, $Y(t) \leftarrow X\lo 2 \rightarrow X\lo 3 \leftarrow X\lo 4 \rightarrow T$, the latter including a collider. Thus, the $d$-separation of $Y(t)$ and $T$ given $A \in \Fs$ is equivalent to that, first, $A$ indexes at least one variable in each of the first three paths, second, $A$ either does not index $X\lo 3$ or indexes at least one of the other nodes in the last path. This means 
\begin{align*}
\As\lo t = \{(1)\} \cup \cone {12} \cup \cone{14}.
\end{align*}
Consequently, the intersection of all the locally minimal sets in $\As\lo t$ is $(1)$, and we have $C\lo t = (3)$ as only $i=3$ satisfies Proposition \ref{prop: what At tells}($i\!i$) with $(1) \in \As\lo t$ and $(13) \not\in \As\lo t$. However, as $i=3$ does not satisfy Proposition \ref{prop: what At tells}($i$), it is kept in the refined $C\lo t$.

%\vspace{0.2cm}

The following graph is the example in Figure \ref{fig: complex} of the main text.

\vspace{-0.25cm}

\begin{figure}[htbp]
	\centering
		\centerline{
			\begin{tikzpicture}
				\node (1) at (-1.25,1) {$X_1$};
				\node (2) at (0,1) {$X_2$};
				\node (3) at (1.25,1) {$X_3$};
				\node (4) at (-1.25,0) {$X_4$};
				\node (5) at (0,0) {$X_5$};
				\node (6) at (1.25,0) {$X_6$};
				\node (7) at (-2.5,0.5) {$Y(t)$};
				\node (8) at (2.25,0.5) {$T$};
    
				\path (1) edge (7);
				\path (1) edge (2);
				\path (2) edge (5);
				\path (2) edge (4);
				\path (3) edge (2);
				\path (3) edge (8);
				\path (4) edge (7);
				\path (6) edge (5);
				\path (6) edge (8);
			\end{tikzpicture}
}
		%\caption{\label{fig3a}(a)}

	\centering
	
\end{figure}

\vspace{-0.25cm}

\noindent
It contains four paths between $Y(t)$ and $T$: $Y(t) \leftarrow X\lo 4 \leftarrow X\lo 2 \leftarrow X\lo 3 \rightarrow T$, $Y(t) \leftarrow X\lo 1 \rightarrow X\lo 2 \leftarrow X\lo 3 \rightarrow T$, $Y(t) \leftarrow X\lo 1 \rightarrow X\lo 2 \rightarrow X\lo 5 \leftarrow X\lo 6 \rightarrow T$,  and $Y(t) \leftarrow X\lo 4 \rightarrow X\lo 2 \rightarrow X\lo 5 \leftarrow X\lo 6 \rightarrow T$, the latter three including colliders. Thus, the $d$-separation of $Y(t)$ and $T$ given $A \in \Fs$ is equivalent to that, first, $A$ indexes at least one variable in the first path, second, $A$ either does not index the collider or indexes at least one of the other nodes in the last three paths. This means 
\begin{align*}
\As\lo t = \cone{12} \cup \cone {14} \cup \cone{23} \cup \cone{36} \cup \{(3), (4), (13), (34), (46), (456)\}.
\end{align*}
Consequently, the intersection of all the locally minimal sets in $\As\lo t$ is the empty set, and we have $C\lo t = (25)$ as these two elements satisfy Proposition \ref{prop: what At tells}($i\!i$) with $(4) \in \As\lo t$, $(24) \not\in \As\lo t$, and $(45) \not\in \As\lo t$. As $i=2$ satisfies Proposition \ref{prop: what At tells}($i$) with $(12)\in \As\lo t$ and $(1) \not\in \As\lo t$ and $i=5$ does not, only $i=2$ is removed from the refined $C\lo t$.

%\vspace{0.2cm}

The following graph is the example in Figure \ref{fig: complex two} of the main text.

\vspace{-0.25cm}

\begin{figure}[htbp]
	\centering
		\centerline{
			\begin{tikzpicture}
				\node (1) at (-0.5,1) {$X_1$};
				\node (2) at (0.8,1) {$X_2$};
				\node (3) at (-0.2,0) {$X_3$};
				\node (4) at (-1.5,0) {$X_4$};
				\node (5) at (1.1,0) {$X_5$};
				\node (6) at (-2.25,1) {$Y(t)$};
				\node (7) at (2,1) {$T$};
    
				\path (1) edge (6);
				\path (1) edge (2);
				\path (2) edge (7);
				\path (3) edge (1);
				\path (4) edge (3);
				\path (5) edge (3);
				\path (4) edge (6);
				\path (5) edge (7);
			\end{tikzpicture}
}
		%\caption{\label{fig3a}(a)}

	\centering
	
\end{figure}

\vspace{-0.25cm}

\noindent
It contains four paths between $Y(t)$ and $T$: $Y(t) \leftarrow X\lo 1 \rightarrow X\lo 2  \rightarrow T$, $Y(t) \leftarrow X\lo 1 \leftarrow X\lo 3 \leftarrow X\lo 5 \rightarrow T$, $Y(t) \leftarrow X\lo 4 \rightarrow X\lo 3 \rightarrow X\lo 1 \rightarrow X\lo 2 \rightarrow T$, and $Y(t) \leftarrow X\lo 4 \rightarrow X\lo 3 \leftarrow X\lo 5 \rightarrow T$, the latter including a collider. Thus, the $d$-separation of $Y(t)$ and $T$ given $A \in \Fs$ is equivalent to that, first, $A$ indexes at least one variable in each of the first three paths, second, $A$ either does not index $X\lo 3$ or indexes at least one of the other nodes in the last path. This means 
\begin{align*}
\As\lo t = \cone {14} \cup \cone{15} \cup \cone{25} \cup \{(1), (12), (234)\}.
\end{align*}
Consequently, the intersection of all the locally minimal sets in $\As\lo t$ is the empty set, and we have $C\lo t = (3)$ as only $i=3$ satisfies Proposition \ref{prop: what At tells}($i\!i$) with $(1) \in \As\lo t$ and $(13) \not\in \As\lo t$. Since $i=3$ also satisfies Proposition \ref{prop: what At tells}($i$) with $(234)\in \As\lo t$ and $(24) \not\in \As\lo t$, it is removed from the refined $C\lo t$.

The following graph is the example in Figure \ref{fig: ct ok cts not sub} of the main text.

\vspace{-0.25cm}

\begin{figure}[htbp]
	\centering
		\centerline{
			
			\begin{tikzpicture}
				\node (1) at (0,1) {$X_1$};
				\node (2) at (-2,0) {$X_2$};
				\node (3) at (0,0) {$X_3$};
				\node (4) at (2,0) {$X_4$};
				\node (5) at (-2,1) {$Y(t)$};
				\node (6) at (2,1) {$T$};
    
				\path (1) edge (5);
				\path (2) edge (5);
				\path (2) edge (3);
				\path (4) edge (3);
				\path (1) edge (3);
				\path (4) edge (6);
				\path (1) edge (6);
			\end{tikzpicture}
}
		%\caption{\label{fig3a}(a)}

	\centering
	
\end{figure}

\vspace{-0.25cm}

\noindent
It contains four paths between $Y(t)$ and $T$: $Y(t) \leftarrow X\lo 1 \rightarrow T$,  $Y(t) \leftarrow X\lo 1 \rightarrow X\lo 3 \leftarrow X\lo 4 \rightarrow T$, $Y(t) \leftarrow X\lo 2 \rightarrow X\lo 3 \leftarrow X\lo 1 \rightarrow T$, $Y(t) \leftarrow X\lo 2 \rightarrow X\lo 3 \leftarrow X\lo 4 \rightarrow T$, the latter including a collider. Thus, the $d$-separation of $Y(t)$ and $T$ given $A \in \Fs$ is equivalent to that, first, $A$ indexes at least one variable in each of the first three paths, second, $A$ either does not index $X\lo 3$ or indexes at least one of the other nodes in the last path. This means 
\begin{align*}
\As\lo t = \{(1)\} \cup \cone {12} \cup \cone{14},
\end{align*}
which is exactly the same as $\As\lo t$ for Figure \ref{fig: ct ok cts not}.

\section{Proof of Theorems and Propositions}
\label{sec: proof}

We now give the proofs of all the theoretical results in the order they appear in the main text.
The proof of Theorem \ref{thm: hd normal x|t} above is placed at the end of this section.

\subsection{Proof of the equivalence property (\ref{as: vs byx btx xa w/o t})}

This equivalence property is important for the proof of Theorem \ref{thm: normal x|t}, so we prove it first. For simplicity, we use $\eta(\cdot)$ to denote the pdf of a continuous random vector. The key to the proof is $T \indep X \mid \beta\lo T\trans X$ by the definition of $\cs T$, which means $T \indep \beta\lo {Y(t)}\trans X \mid (\beta\lo T\trans X, X\lo A)$ for any $A\in \Fs$. Under the common support condition (2), this means
\begin{align}\label{pf: cond equiv t}
\eta(\beta\lo {Y(t)}\trans X \mid \beta\lo T\trans X, X\lo A, T) = \eta(\beta\lo {Y(t)}\trans X \mid \beta\lo T\trans X, X\lo A)
\end{align}
for any $A \in \Fs$. If $\beta\lo {Y(t)}\trans X \indep \beta\lo T\trans X | X\lo A$, then $\eta(\beta\lo {Y(t)}\trans X | \beta\lo T\trans X, X\lo A)$ reduces to $\eta(\beta\lo {Y(t)}\trans X | X\lo A)$, which, by (\ref{pf: cond equiv t}), means $\eta(\beta\lo {Y(t)}\trans X \mid \beta\lo T\trans X, X\lo A, T) = \eta(\beta\lo {Y(t)}\trans X \mid X\lo A)$ and consequently
\begin{align}\label{pf: cond equiv t 2}
\eta(\beta\lo {Y(t)}\trans X \mid \beta\lo T\trans X, X\lo A, T) = \eta(\beta\lo {Y(t)}\trans X \mid X\lo A, T).
\end{align}
The latter readily implies $\beta\lo {Y(t)}\trans X \indep \beta\lo T\trans X \mid (X\lo A, T)$. Conversely, if $\beta\lo {Y(t)}\trans X \indep \beta\lo T\trans X \mid (X\lo A, T)$, then as is (\ref{pf: cond equiv t 2}). By (\ref{pf: cond equiv t}), we have
\begin{align*}
\eta(\beta\lo {Y(t)}\trans X \mid \beta\lo T\trans X, X\lo A) = \eta(\beta\lo {Y(t)}\trans X \mid X\lo A, T),
\end{align*}
which means that $\eta(\beta\lo {Y(t)}\trans X \mid X\lo A, T)$ is invariant of $T$ and thus reduces to $\eta(\beta\lo {Y(t)}\trans X \mid X\lo A)$, and consequently $\eta(\beta\lo {Y(t)}\trans X \mid \beta\lo T\trans X, X\lo A) = \eta(\beta\lo {Y(t)}\trans X \mid X\lo A)$. The latter means $\beta\lo {Y(t)}\trans X \indep \beta\lo T\trans X \mid X\lo A$. This completes the proof.

\subsection{Proof of Theorem \ref{thm: normal x|t}}

For simplicity, we omit the phrase ``almost surely" in this proof. Under the ignorability assumption (1), $X\lo A$ satisfies (3) if and only if, for any $h(Y(t))$ with finite $E\{ h\udex 2 (Y(t)) | X \}$,
\begin{align}\label{eq: prf thm1 1}
E\{h(Y(t))\mid X\lo A\} E\{T \mid X\lo A\} = E[E\{h(Y(t))|X\} E \{T|X\} \mid X\lo A].
\end{align}
This is because $T$ is binary and the right-hand side of (\ref{eq: prf thm1 1}) equals $E[\{h(Y(t)) T | X\}X\lo A]$ or equivalently $E\{h(Y(t)) T|X\lo A\}$ under (1). By the definition of $\beta\lo {Y(t)}$ and $\beta\lo T$, (\ref{eq: prf thm1 1}) is equivalent to
\begin{align*}
E\left[E\{h(Y(t)) | \beta\lo {Y(t)}\trans X\} | X\lo A\right] E\left[E\{T| \beta\lo T\trans X\} | X\lo A\right] = E\left[E\{h(Y(t))|\beta\lo {Y(t)}\trans X\} E \{T|\beta\lo T\trans X\} | X\lo A \right],
\end{align*}
which is equivalent to $\cov[E\{h(Y(t))\mid \beta\lo {Y(t)}\trans X\}, E(T \mid \beta\lo T\trans X) \mid X\lo A] = 0$. Under Assumption 1, this is also equivalent to $\beta\lo {Y(t)}\trans X \indep \beta\lo T\trans X \mid X\lo A$, which, under the common support condition (2) and by (14), is further equivalent to 
\begin{align}\label{eq: prf thm1 4}
\beta\lo {Y(t)}\trans X \indep \beta\lo T\trans X \mid (X\lo A, T). 
\end{align}
Under Assumption 2, (\ref{eq: prf thm1 4}) is equivalent to $\cov (\beta\lo {Y(t)}\trans X, \beta\lo T\trans X \mid X\lo A, T) = 0$, which, by simple algebra, can be rewritten as (13). This completes the proof.

\subsection{Proof of Theorem \ref{thm: hf=f normal x|t}}

By simple algebra, $\widehat \Sigma\udex {(s)}$ is a $n\udex {1/2}$-consistent estimator of $\Sigma\udex {(s)}$ for $s=0, 1$. From Section \ref{sec: review}, we also have $\|\widehat \M\lo {Y(t)} - \M\lo {Y(t)}\| = \bop (n\udex {-1/2})$ and $\|\widehat \M\lo T - \M\lo T \| = \bop (n\udex {-1/2})$. These readily imply 
\begin{align}\label{eq: prf thm2 1}
\max \{ |\widehat f\lo t (A) - f\lo t (A) |: A \in \Fs\} = \bop (n\udex {-1/2}). 
\end{align}
Also see the proof of Theorem \ref{thm: hd normal x|t} below for more relative details. Suppose $\As\lo t \neq \Fs$ first. Since $\As\lo t = \{A \in \Fs: f\lo t(A) = 0\}$, (\ref{eq: prf thm2 1}) implies that, for $c\lo \min \equiv \min\{f\lo t(A): A \not\in \As\lo t\} > 0$,
\begin{align}\label{eq: prf thm2 2}
\max\{\widehat f\lo t (A): A \in \As\lo t\} = \bop (n\udex {-1/2}) \mbox{ and } \min\{\widehat f\lo t (A): A \not\in \As\lo t\} = c\lo \min + \bop (n\udex {-1/2}), 
\end{align}
which means $\max\{\widehat f\lo t (A): A \in \As\lo t\} < \min\{\widehat f\lo t (A): A \not\in \As\lo t\}
$ or equivalently 
\begin{align}\label{eq: prf thm2 3}
\As\lo t = \{A\lo k\udex {(t)}: k > k\lo 0\udex {(t)}\}
\end{align}
for some $k\lo 0\udex {(t)} \in \{1,\ldots, 2\udex p\}$ with probability converging to one. Thus, it suffices to prove $\tau(t) = k\lo 0\udex {(t)}$ with probability converging to one. Given (\ref{eq: prf thm2 3}), (\ref{eq: prf thm2 2}) means 
\begin{align*}
\max\{\widehat f\lo t (A\lo k\udex {(t)}): k > k\lo 0\udex {(t)}\} = \bop (n\udex {-1/2}) \mbox{ and }
\min\{\widehat f\lo t (A\lo k\udex {(t)}): k \leq k\lo 0\udex {(t)}\} = c\lo \min + \sop (n\udex {-1/2}).
\end{align*}
Together with $c\lo n =o(1)$ and $n\udex {-1/2}/c\lo n = o(1)$, we have, uniformly for $k\in \{1,\ldots, 2\udex p-1\}$,
\begin{align*}
R\lo t(k) = \frac {\widehat f\lo t(A\lo {k+1}\udex {(t)}) + c\lo n} {\widehat f\lo t(A\lo {k}\udex {(t)}) + c\lo n} = \left\{
\begin{matrix} 
c\lo \min /c\lo \max + c\lo k + \bop (c\lo n) & \mbox{ if } k < k\lo 0\udex {(t)} \\
\bop (c\lo n) & \mbox{ if } k = k\lo 0\udex {(t)} \\
1 + \bop (n\udex {-1/2}/c\lo n) & \mbox{ if } k > k\lo 0\udex {(t)}
\end{matrix}\right.   
\end{align*}
for $c\lo \max \equiv \max\{f\lo t(A): A \not\in \As\lo t\}$ and some $c\lo k \geq 0$. Together with $R\lo t(0) = c\lo 0 >0$, this readily implies $\tau(t) = k\lo 0\udex {(t)}$ with probability converging to one. Now suppose $\As\lo t = \Fs$, then the first part of (\ref{eq: prf thm2 2}) implies $\min\{R\lo t(k):k=1,\ldots, 2\udex p-1\} = 1 + \bop(n\udex {-1/2}/c\lo n)$. Together with $R\lo t(0) = c\lo 0 <1$, we have $\tau(t)=0$ or equivalently $\widehat \As\lo t = \Fs$ with probability converging to one.  This completes the proof.

\subsection{Proof of Theorem \ref{thm: hf=f gcopula x|t}}

The proof follows the proof of Theorem \ref{thm: hf=f normal x|t} if we show $\|\widehat \M\lo {Y(t)|\zeta(X)} -\M\lo {Y(t)|\zeta(X)}\| = \bop (n\udex {-1/2})$, $\|\widehat \M\lo {T|\zeta(X)} -\M\lo {T|\zeta(X)}\| = \bop (n\udex {-1/2})$, and $\|\widehat \Omega\udex {(s)} - \Omega\udex {(s)}\| = \bop (n\udex {-1/2})$ for $s=0, 1$, which would be obvious if we replace $\widehat \zeta\lo {i}(\cdot)$ with the true $\zeta\lo i(\cdot)$ for $i=1,\ldots, p$. Thus, it suffices to show that the impact of using $\widehat \zeta\lo {i}(\cdot)$ in place of $\zeta\lo i(\cdot)$ is $\bop (n\udex {-1/2})$ in estimating the conditional and marginal moments in $\widehat \M\lo {Y(t)}$, $\widehat \M\lo T$, and $\widehat \Omega\udex {(s)}$. These resemble the asymptotic studies for the normal-score estimator in the Gaussian copula model \citep{klaassen1997,serfling2009,hoff2014,mai2023}. For example, the $n\udex {1/2}$-consistency of $\widehat \Omega\udex {(s)}$ can be proved exactly the same as for Theorem $3.1$ of \cite{klaassen1997}, and, if we use the sliced inverse regression for $\M\lo {Y(t)}$ and assume that $\zeta\lo i(X\lo i)$ has the standard normal distribution without loss of generality, then, similarly to the proof of Theorem $3.1$ in \cite{klaassen1997}, we have 
\begin{align*}
& E\lo n [\{\widehat \zeta\lo i (X\lo i) - \zeta\lo i (X\lo i)\} \delta (Y\lo S(t)=h, T=t) ]
= E\lo n [\{\widehat \zeta\lo {i, t} (X\lo i) - \zeta\lo i (X\lo i)\} \delta (Y\lo S(t)=h, T=t) ] \\
& \hspace{0.5cm} = E\lo n \left[ [\Phi\inv \{\widehat F\lo {i, t}\udex *(X\lo i)\} - \Phi\inv \{F\lo {i, t}(X\lo i)\}] \delta (Y\lo S(t)=h, T=t) \right] \\
& \hspace{0.5cm} = E\lo n [ \{\widehat F\lo {i, t}\udex *(X\lo i) - F\lo {i, t}(X\lo i)\} \delta (Y\lo S(t)=h, T=t) / \Phi\{F\lo {i,t} (X\lo i)\} ] + \sop(n\udex {-1/2}) \\
& \hspace{0.5cm} = E [ \{\widehat F\lo {i, t}\udex *(X\lo i) - F\lo {i, t}(X\lo i)\} \delta (Y\lo S(t)=h, T=t) / \Phi\{F\lo {i,t} (X\lo i)\} ] + \sop(n\udex {-1/2}) = \bop (n\udex {-1/2}),
\end{align*}
where the last equality is by Donsker's Theorem. The $n\udex {1/2}$-consistency of $\widehat \M\lo {T|\zeta(X)}$ involves the merged estimator $\widehat \zeta\lo i(\cdot)$ and thus is more tedious, but the proof essentially repeats those above as $\widehat \zeta\lo i(\cdot)$ has a simple form and the coefficients $(\widehat a\lo i, \widehat b\lo i)$ therein are derived by the simple minimal truncated squared loss. Thus, we omit the details. This completes the proof.

\subsection{Proof of Proposition \ref{prop: global min set}}

{\it Proof of ($i$)} Since $A\lo {Y(t)} \cap A\lo T$ indexes the variables of $X$ that are uniquely informative to both $Y(t)$ and $T$, each of which forms a fork between $Y(t)$ and $T$, $A\lo {Y(t)} \cap A\lo T$ must be included in each locally minimal $A \in \As\lo t$ for the conditional $d$-separation of $Y(t)$ and $T$ given $X\lo A$. Conversely, if $i \not\in A\lo {Y(t)} \cap A\lo T$, that is, $X\lo i$ does not form a fork between $Y(t)$ and $T$, then the rest of $X$ still induces the conditional $d$-separation of $Y(t)$ and $T$ and consequently satisfies (3), which means that there exists $A \in \As\lo t$ with $i\not\in A$. Thus, the intersection of all $A \in \As\lo t$ must be a subset of $A\lo {Y(t)} \cap A\lo T$. These together imply that $A\lo {Y(t)} \cap A\lo T$ is exactly the intersection of all the locally minimal sets in $\As\lo t$.

\vspace{0.2cm}

{\it Proof of ($i\!i$)} Suppose $A \in \As\lo t$ includes some $i \not\in A\lo {Y(t)} \cap A\lo T$. From the proof of ($i$) above, there must exist some $B \in \As\lo t$ with $i \not\in B$, which means $A \not\subseteq B$. Thus, if there exists the unique minimal set in $\As\lo t$, this set must be a subset of $A\lo {Y(t)} \cap A\lo T$. Together with ($i$) above, this set must further be $A\lo {Y(t)} \cap A\lo T$, which also means $A\lo {Y(t)} \cap A\lo T \in \As\lo t$. The ``only if" part is obvious based on ($i$) above. This completes the proof. 
%say $i \not\in A\lo {T}$ without loss of generality. Let $X\lo {-i}$ be the rest of $X$ after removing $X\lo i$. We have $T\indep X \mid X\lo {-i}$. Under the ignorability assumption (1), this implies $Y(t) \indep T \mid X\lo {-i}$, which means that $\{i\}\udex C$, which denotes $\{j=1,\ldots, p: j\neq i\}$, is included in $\As\lo t$. Thus, there exists a locally minimal set in $\As\lo t$ that is a subset of $\{i\}\udex C$, which means it differs with $A$. 
%Under Assumption \ref{as: vs xy xt xa}, any $A \in \As$ must satisfy $X_{A_{Y(t)}} \indep X_{A_T} \mid X\lo A$, which means $X_{A_{Y(t)} \cap A_T} \indep X_{A_{Y(t)} \cap A_T} \mid X\lo A$ and consequently $A\lo {Y(t)} \cap A\lo T \subseteq A$. 
%The coincidence between $A\lo 0$ and $A\lo {Y(t)} \cap A\lo T$ under the unique existence of $A\lo 0$ and Assumption \ref{as: vs xy xt xa} is an immediate corollary of ($i$) and ($i\!i$), so we omit its proof. This completes the proof. 

\subsection{Proof of Proposition \ref{prop: what At tells}}

The proof is a direct application of the definition of $d$-separation (see Appendix \ref{sec: dag} above).

\vspace{0.2cm}

{\it Proof of ($i$)-part one} For $i=1,\ldots, p$, if there exists $A \in \As\lo t$ such that $A \backslash \{i\} \not\in \As\lo t$, then there exists a path between $Y(t)$ and $T$ that is blocked by $A$ but not blocked by $A \backslash \{i\}$. Thus, this path either does not include any collider, under which it exactly includes $X\lo i$ among all indexed by $A$, or it includes a set of colliders and its non-colliders exactly include $X\lo i$ among all indexed by $A$. In both cases, $X\lo i$ is a non-collider in some path between $Y(t)$ and $T$. 

\vspace{0.2cm}

{\it Proof of ($i$)-part two} First, any $A \in \Ns\lo t$ means $A$ includes at least one non-collider in every path between $Y(t)$ and $T$ that has colliders; otherwise, we can add to $A$ exactly all the colliders in the path where $A$ does not include any non-collider, and the resulting $B \supset A$ does not block $Y(t)$ and $T$ in that path and thus is not in $\As\lo t$. Conversely, we have $A \in \Ns\lo t$ as long as $A$ includes at least one non-collider in every path between $Y(t)$ and $T$; the proof is easy from the definition of $d$-separation and omitted. Now suppose $A \in \Ns\lo t$, $A \backslash \{i\} \in \As\lo t$, and $A \backslash \{i\} \not\in \Ns\lo t$. Then $A \backslash \{i\}$ does not include any non-collider in some path that has colliders, but $A$ does include a non-collider in the same path. Thus, $X\lo i$ must be the non-collider in that path.  

\vspace{0.2cm}

{\it Proof of ($i\!i$)} Continued from the proof of ($i$)-part two above, if $A \in \As\lo t$ and there exists $B \in \Fs$ such that $A \cup B \not\in \As\lo t$ and  $A\cup C \in \As\lo t$ for any $C\subset B$, then $A$ does not include any non-collider in at least one path that has colliders, $B$ includes all the colliders in at least one of these paths, and any $C \subset B$ must not include all the colliders in any of these paths. These together imply that $B$ exactly consists of all the colliders in the same path.

\subsection{Proof of Theorem \ref{thm: hd normal x|t}}

\def\N{\mathrm {N}}
\def\one{\mathrm {I}}
\def\two{\mathrm {I\!I}}

The proof follows the proof of Theorem \ref{thm: hf=f normal x|t} as long as we can show:

\noindent
($i$) $\max\{|\widehat f\lo t(A) - f\lo t(A)|: A \in \Fs\} = \bop (r\lo {n,p})$;

\noindent
($i\!i$) if $\As\lo t\neq \Fs$, then $1/\min \{f\lo t(A): A \not\in \As\lo t\} =O(1)$;

\noindent
($i\!i\!i$) if $\As\lo t\neq \Fs$, then $\max \{f\lo t(A): A \not\in \As\lo t\} = O(1)$.

\vspace{0.2cm}

\noindent
Proof of $($i$)$: for any $A \in \Fs$, we can write $f\lo t(A)$ as $\sum\lo {s=0,1} \|\M\lo {Y(t)}\trans\{\Sigma\udex {(s)} P(\Sigma\udex {(s)}, I\lo A) - \Sigma\udex {(s)}\}\M\lo T\|$, and likewise write $\widehat f\lo t(A)$. Since $\|\widehat \M\lo {Y(t)} - \M\lo {Y(t)}\| = \bop (r\lo {n,p})$ and $r\lo {n,p}=o(1)$, under Assumption \ref{as: signal strength}($i$) that $\M\lo {Y(t)} \in \Ms$, we have $\|\widehat \M \lo {Y(t)}\| = \bop (1)$. Similarly, we also have $\|\widehat \M \lo {T}\| = \bop (1)$. Recall that, for general matrices $\M, \N$ and their estimators $\widehat \M, \widehat \N$,
\begin{align*}
| \|\widehat \M \widehat \N\| - \|\M \N\| | \leq \|\widehat \M \widehat \N - \M \N\| = \|\widehat \M \widehat \N - \widehat \M \N + \widehat \M \N -  \M \N\| \\
\leq \|\widehat \M \widehat \N - \widehat \M \N \| + \|\widehat \M \N -  \M \N\| 
\leq \|\widehat \M \| \|\widehat \N - \N \| + \|\widehat \M -  \M \| \|\N\|,
\end{align*}
where the first two inequalities are the triangle inequalities for norms. These together imply
\begin{align*}
& \max\{|\widehat f\lo t(A) - f\lo t(A)|: A \in \Fs\} \\
& \hspace{0.3cm} 
\leq \sum_ {s=0,1} \left[\|\widehat \M\lo {Y(t)}- \M\lo {Y(t)}\| \max\{\|\widehat \Sigma\udex {(s)} P(\widehat \Sigma\udex {(s)}, I\lo A) - \widehat \Sigma\udex {(s)}\|: A \in \Fs\} \|\widehat \M\lo T\| \right.\\
& \hspace{1.1cm}
+ \|\M\lo {Y(t)}\| \max\{\|\widehat \Sigma\udex {(s)} P(\widehat \Sigma\udex {(s)}, I\lo A) - \widehat \Sigma\udex {(s)} - \Sigma\udex {(s)} P(\Sigma\udex {(s)}, I\lo A) +\Sigma\udex {(s)}\| : A \in \Fs\} \|\widehat \M\lo T\| \\
& \hspace{1.1cm}
+ \left.\|\M\lo {Y(t)}\| \max\{\|\Sigma\udex {(s)} P(\Sigma\udex {(s)}, I\lo A) - \Sigma\udex {(s)}\|: A \in \Fs\} \|\widehat \M\lo T - \M\lo T\| \right] \\
& \hspace{0.3cm}
= \bop (r\lo {n,p}) \max\{\|\widehat \Sigma\udex {(s)} P(\widehat \Sigma\udex {(s)}, I\lo A) - \widehat \Sigma\udex {(s)}\| + \|\Sigma\udex {(s)} P(\Sigma\udex {(s)}, I\lo A) - \Sigma\udex {(s)}\|: A \in \Fs\} \\
& \hspace{1.1cm}
+ \bop (1) \max\{\|\widehat \Sigma\udex {(s)} P(\widehat \Sigma\udex {(s)}, I\lo A) - \widehat \Sigma\udex {(s)} - \Sigma\udex {(s)} P(\Sigma\udex {(s)}, I\lo A) +\Sigma\udex {(s)}\| : A \in \Fs\} \\
& \hspace{0.3cm} \equiv \epsilon\lo 1 + \epsilon\lo 2.
\end{align*}
Since $I\lo p - P(\widehat \Sigma\udex {(s)}, I\lo A)$ is a projection matrix under the inner product $\langle u, v\rangle = u\trans \widehat \Sigma\udex {(s)} v$, we have $\|\widehat \Sigma\udex {(s)} - \widehat \Sigma\udex {(s)} P(\widehat \Sigma\udex {(s)}, I\lo A)\| = \|\widehat \Sigma\udex {(s)} \{I\lo p - P(\widehat \Sigma\udex {(s)}, I\lo A)\}\| \leq \|\widehat \Sigma\udex {(s)}\|$ for all $A \in \Fs$, which means 
\begin{align*}
\max\{\|\widehat \Sigma\udex {(s)} P(\widehat \Sigma\udex {(s)}, I\lo A) - \widehat \Sigma\udex {(s)}\|: A \in \Fs\} \leq \|\widehat \Sigma\udex {(s)}\| = \bop (1).
\end{align*}
Likewise, we also have $\max\{\| \Sigma\udex {(s)} P( \Sigma\udex {(s)}, I\lo A) - \Sigma\udex {(s)}\|: A \in \Fs\} = O(1)$, which together imply $\epsilon\lo 1 = \bop (r\lo {n,p})$. Since $\widehat \Sigma\udex {(s)} P(\widehat \Sigma\udex {(s)}, I\lo A) = \widehat \Sigma\lo {\cdot, A}\udex {(s)} (\widehat \Sigma\lo {A,A}\udex {(s)})\inv \widehat \Sigma\lo {A,\cdot}\udex {(s)}$, we have
\begin{align}\label{eq: prf thm4 1}
& \|\widehat \Sigma\udex {(s)} P(\widehat \Sigma\udex {(s)}, I\lo A) - \Sigma\udex {(s)} P(\Sigma\udex {(s)}, I\lo A)\| 
= \|\widehat \Sigma\lo {\cdot, A}\udex {(s)} (\widehat \Sigma\lo {A,A}\udex {(s)})\inv \widehat \Sigma\lo {A,\cdot}\udex {(s)} - \Sigma\lo {\cdot, A}\udex {(s)} (\Sigma\lo {A,A}\udex {(s)})\inv \Sigma\lo {A,\cdot}\udex {(s)}\| \nonumber  \\
& \hspace{0.8cm} \leq \|\widehat \Sigma\lo {\cdot, A}\udex {(s)} - \Sigma\lo {\cdot, A}\udex {(s)}\| \|(\widehat \Sigma\lo {A,A}\udex {(s)})\inv \| \|\widehat \Sigma\lo {A,\cdot}\udex {(s)}\| \nonumber + \| \Sigma\lo {\cdot, A}\udex {(s)}\| \|(\widehat \Sigma\lo {A,A}\udex {(s)})\inv \|\Sigma\lo {A,A}\udex {(s)} - \widehat \Sigma\lo {A,A}\udex {(s)}\| \|(\Sigma\lo {A,A}\udex {(s)})\inv \| \|\widehat \Sigma\lo {A,\cdot}\udex {(s)}\| \nonumber \\
& \hspace{1.2cm} + \| \Sigma\lo {\cdot, A}\udex {(s)}\| \|(\Sigma\lo {A,A}\udex {(s)})\inv\| \|\widehat \Sigma\lo {A,\cdot}\udex {(s)} - \Sigma\lo {A,\cdot}\udex {(s)}\|
\end{align}
Since $\|\N\| \leq \|\M\|$ for a general matrix $\M$ and its arbitrary submatrix $\N$, we have  
\begin{align*}
\|\widehat \Sigma\lo {A,\cdot}\udex {(s)}\| \leq \|\widehat \Sigma\udex {(s)}\| = \bop (1), \,\, \|\Sigma\lo {A,\cdot}\udex {(s)}\| \leq \|\Sigma\udex {(s)}\| = O (1), \,\, \|\widehat \Sigma\lo {A, \cdot} - \Sigma\lo {A, \cdot}\| \leq \|\widehat \Sigma - \Sigma \| = \bop (r\lo {n,p})    
\end{align*} 
uniformly for $A \in \Fs$. Since $\|\Sigma\inv\| = O(1)$ under Assumption \ref{as: signal strength}($i$), we also have 
\begin{align*}
\|(\Sigma\lo {A,A}\udex {(s)})\inv\| \leq \|\{\Sigma\lo {A,A}\udex {(s)} - \Sigma\lo {A,-A}\udex {(s)} (\Sigma\lo {-A,-A}\udex {(s)})\inv \Sigma\lo {-A, A}\udex {(s)}\}\inv\| = \|(\Sigma\udex {(s)})\inv\lo {A,A}\| \leq \|(\Sigma\udex {(s)})\inv\| = O(1), 
\end{align*}
and similarly $\|(\widehat \Sigma\lo {A,A}\udex {(s)})\inv\| \leq \|(\widehat \Sigma\udex {(s)})\inv\| = \{\lambda\lo \min(\Sigma\udex {(s)}) + \bop (r\lo {n,p})\}\inv = \bop (1)$ uniformly for $A \in \Fs$, where the first equality is by Weyl's Theorem. Plugging these into (\ref{eq: prf thm4 1}), we have, uniformly for $A \in \Fs$, $\|\widehat \Sigma\udex {(s)} P(\widehat \Sigma\udex {(s)}, I\lo A) - \Sigma\udex {(s)} P(\Sigma\udex {(s)}, I\lo A)\| = \bop (r\lo {n,p})$, which implies
\begin{align*}
\epsilon\lo 2 = & \bop (1) \left[ \|\widehat \Sigma\udex {(s)} - \Sigma\udex {(s)}\| + \max\{\|\widehat \Sigma\udex {(s)} P(\widehat \Sigma\udex {(s)}, I\lo A) - \Sigma\udex {(s)} P(\Sigma\udex {(s)}, I\lo A)\|: A \in \Fs\} \right]
= \bop (r\lo {n,p}).
\end{align*}
Together with $\epsilon\lo 1 = \bop (r\lo {n,p})$ shown above, this implies ($i$).

\medskip

\noindent 
Proof of ($i\!i$) Write $\M\lo {Y(t)} = \beta\lo {Y(t)} \N\lo {Y(t)}$ and $\M\lo T = \beta\lo T \N\lo T$ where both $\N\lo {Y(t)}$ and $\N\lo T$ have full row rank. Since $\M\lo {Y(t)}, \M\lo T \in \Ms$ and $\beta\lo {Y(t)}$ and $\beta\lo T$ are sequences of semi-orthogonal matrices, we have $\lambda\lo \min\inv (\N\lo {Y(t)}) = O(1)$ and $\lambda\lo \min\inv (\N\lo {T}) = O(1)$. Together with Assumption 2 and Assumption \ref{as: signal strength}($i\!i$), the former again indicating the constant variance condition (10) on $X|T$, these imply
\begin{align}\label{eq: prf thm4 2}
& \min\{f\lo t(A): A \not\in \As\} 
= \min\{\sum\lo {s=0,1} \|\N\lo {Y(t)}\trans \beta\lo {Y(t)}\trans \{\Sigma\udex {(s)} P(\Sigma\udex {(s)}, I\lo A) - \Sigma\udex {(s)}\} \beta\lo T \N\lo T\| : A \not\in \As\} \nonumber \\
& \hspace{0.6cm} \geq \min\{\sum\lo {s=0,1} \lambda\lo \min (\N\lo {Y(t)}) \lambda\lo \min( \N\lo T) \|\beta\lo {Y(t)}\trans \{\Sigma\udex {(s)} P(\Sigma\udex {(s)}, I\lo A) - \Sigma\udex {(s)}\} \beta\lo T\| : A \not\in \As\} \nonumber \\
& \hspace{0.6cm} = \lambda\lo \min (\N\lo {Y(t)}) \lambda\lo \min( \N\lo T) \min\{\sum\lo {s=0,1} \|\cov(\beta\lo {Y(t)}\trans X, \beta\lo T\trans X | X\lo A, T=s)\| : A \not\in \As\} > c\udex *
\end{align}
for some $c\udex * > 0$. This readily implies ($i\!i$).

\medskip

\noindent 
Proof of ($i\!i\!i$) Similar to (\ref{eq: prf thm4 2}), we have
\begin{align*}
\max\{f\lo t(A): A \not\in \As\}  \leq \|\N\lo {Y(t)}\| \|\N\lo T\| \max\{\sum\lo {s=0,1} \|\cov(\beta\lo {Y(t)}\trans X, \beta\lo T\trans X | X\lo A, T=s)\| : A \not\in \As\}.
\end{align*}
Under Assumption 2 and Assumption \ref{as: signal strength}($i$), we also have $\|\N\lo {Y(t)}\| = O(1)$, $\|\N\lo T\| = O(1)$, and 
\begin{align*}
\|\cov(\beta\lo {Y(t)}\trans X, \beta\lo T\trans X | X\lo A, T=s)\| 
& = \| \beta\lo {Y(t)}\trans \{\Sigma\udex {(s)} - \Sigma\udex {(s)} P(\Sigma\udex {(s)}, I\lo A) \} \beta\lo {T}\| \\
& \leq \|\Sigma\udex {(s)} - \Sigma\udex {(s)} P(\Sigma\udex {(s)}, I\lo A)\| \leq \|\Sigma\udex {(s)}\| = O(1)
\end{align*}
uniformly for $A \in \Fs$. These imply $\max\{f\lo t(A): A \not\in \As\} = O(1)$. This completes the proof.
%\begin{align*}
%& \max\{f\lo t(A): A \not\in \As\} \, / \, \min\{f\lo t(A): A \not\in \As\} \\
%& \hspace{0.3cm} = \frac{\|\N\lo {Y(t)}\| \|\N\lo T\|}{\lambda\lo \min(\N\lo {Y(t)}) \lambda\lo \min (\N\lo T)} \left[\min\{\sum\lo {s=0,1} \|\cov(\beta\lo {Y(t)}\trans X, \beta\lo T\trans X | X\lo A, T=s)\| : A \not\in \As\} \right]\inv.
%\end{align*}

\section{Additional simulation results for $p=20$}

We now raise $p$ to $20$ for the simulation models in Section \ref{sec: sim} of the main text, with $n$ still set at $400$ and $800$ sequentially. For clarity, we change $X\lo {4,8,9,10}\sim N(0, .6I\lo 4)$ to $X\lo {4,8,\ldots,20} \sim N(0, .6I\lo{14})$ in Model $3$; the other model settings remain the same. The results based on $2000$ independent runs are summarized in Table \ref{table: sim p=20}. Compared with Table 1 in the main text for $p=10$, most phenomena can still be observed in Table \ref{table: sim p=20}, indicating the overall consistency of the proposed methods as well as their robustness against the dimensionality of $X$. The only notable step-down is that, when $n=400$, the Gaussian copula estimator only selects $52\%$ of the sufficient adjustment sets in $\As\lo 1$ and $39\%$ of those in $\As\lo 0$ in Model 5, where the weak signals are again present. In addition, the slightly larger $\rho\lo t$ and $\pi\lo t$ and the slightly smaller $\omega\lo t$ for $t=0, 1$ in Models 1-2 compared with Table 1 suggest that $c\lo n = .2 n\udex {-1/2} \log n$ becomes a slightly more conservative choice as $p$ grows, that is, it leads to a larger $\widehat \As\lo t$ that includes more sets in $\Fs$.

\def\Ts{{\mathcal T}}
\def\Fs{{\mathcal F}}

\begin{table}[b]
\vspace*{-6pt}
\centering
\def\~{\hphantom{0}}
\begin{tabular*}{\columnwidth} {@{}l@{\extracolsep{\fill}}r@{\extracolsep{\fill}}r@{\extracolsep{\fill}}r@{\extracolsep{\fill}}r@{\extracolsep{\fill}}r@{\extracolsep{\fill}}r@{\extracolsep{\fill}}r@{\extracolsep{\fill}}r@{\extracolsep{\fill}}r@{\extracolsep{\fill}}r@{\extracolsep{\fill}}r@{}} \\
	\hline
\empty & \multicolumn{2}{c}{{Model 1}} & \multicolumn{2}{c}{{Model 2}} & \multicolumn{2}{c}{{Model 3}} & \multicolumn{2}{c}{{Model 4}} & \multicolumn{2}{c}{{Model 5}}  \\ \hline 
$n=400$ &\multicolumn{1}{c}{MN}  & \multicolumn{1}{c}{GC} & \multicolumn{1}{c}{MN}  & \multicolumn{1}{c}{GC}  & \multicolumn{1}{c}{MN} & \multicolumn{1}{c}{GC} & \multicolumn{1}{c}{MN}  & \multicolumn{1}{c}{GC} & \multicolumn{1}{c}{MN} & \multicolumn{1}{c}{GC}   \\ \hline
 %\hline \\ [-6pt]
	%\hline 
	$\rho\lo 0/\omega\lo 0$ & 94/89&	94/89&	90/93&	89/93&	100/74&	100/74&	17/79&	87/69	&16/80&	52/70 \\
	$\rho\lo 1/\omega\lo 1$ &  94/89&	93/89&	89/90&	88/90&	92/76&	94/76&	16/80&	82/65	&16/80&	39/76 \\		
	$\pi\lo 0 / \pi\lo 1$ & 91/92&	91/91&	86/86&	85/84&	100/92&	100/94&	6/5&	87/82&	5/4&	52/39
\\
	$\Ts\lo 0 / \Fs\lo 0$ & 34/81&	33/83&	49/58&	47/48&	$-/0$&	$-/0$&	$-/0$&	$-/1$&	$-/0$&	$-/0$\\
	$\Ts\lo 1 / \Fs\lo 1$ & 33/88&	32/86&	46/107&	46/107&	$-/6$&	$-/5$&	$-/0$&	$-/3$&	$-/0$&	$-/0$\\ \hline
$n=800$ &\multicolumn{1}{c}{MN}  & \multicolumn{1}{c}{GC} & \multicolumn{1}{c}{MN}  & \multicolumn{1}{c}{GC}  & \multicolumn{1}{c}{MN} & \multicolumn{1}{c}{GC} & \multicolumn{1}{c}{MN}  & \multicolumn{1}{c}{GC} & \multicolumn{1}{c}{MN} & \multicolumn{1}{c}{GC}   \\ \hline
$\rho\lo 0/\omega\lo 0$ &99/96&	99/96&	99/98&	99/98&	100/74&	100/74&	21/76&	100/97&	21/75&	100/80 \\
	$\rho\lo 1/\omega\lo 1$ &  99/96&	99/96&	98/96&	98/96&	100/74&	100/74	&22/74&	100/94&	22/74&	98/74\\		
	$\pi\lo 0 /\pi\lo 1$ & 97/97&	97/97&	95/94&	96/94&	100/100&	100/100&	8/9&	100/100&	8/8&	100/98\\
	$\Ts\lo 0 / \Fs\lo 0$ & 74/19&	73/19&	83/11&	84/12&	$-/0$&	$-/0$&	$-/0$&	$-/0$&	$-/0$&	$-/0$
 \\
$\Ts\lo 1 / \Fs\lo 1$ &  74/18&	74/19&	80/49&	80/50&	$-/5$&	$-/2$&	$-/0$&	$-/0$&	$-/0$&	$-/0$ \\ \hline
 %\hline
\end{tabular*}%\vskip18pt
\caption{The performance of the proposed methods when $p=20$: the notations and the meanings of numbers in each cell follow those in Table \ref{table: sim} of the main text.}
\label{table: sim p=20}
\end{table}

\vskip 0.2in
\bibliography{sample}

\end{document}